\documentclass[preprint,onecolumn,superscriptaddress,amsfonts,amssymb]{revtex4-2}
\usepackage{graphicx}
\usepackage{dcolumn}
\usepackage{xcolor}
\usepackage{soul}
\usepackage{color} 
\usepackage{bm}
\usepackage{times}
\usepackage{natbib}
\usepackage{amsmath}
\usepackage{amssymb}
\usepackage[normalem]{ulem}
\usepackage[none]{hyphenat}
\usepackage{siunitx}

\begin{document}

\newcommand{\de }{$^{\circ}$}
\newcommand{\JL}[1]{\textcolor{blue}{{\bf JL: #1 }}} 

\newcommand{\out}[1]{{\color{red}\sout{#1}}}
\newcommand{\add}[1]{{\color{blue}{#1}}}
\newcommand{\com}[1]{{\color{magenta} #1}}

\title{Ferromagnetism on an atom-thick \& extended 2D-metal-organic framework}

\author{Jorge Lobo-Checa} \email{jorge.lobo@csic.es}
\affiliation{Instituto de Nanociencia y Materiales de  Arag\'on (INMA), CSIC-Universidad de Zaragoza, Zaragoza 50009, Spain}
\affiliation {Departamento de F\'{\i}sica de la Materia Condensada, Universidad de Zaragoza, E-50009 Zaragoza, Spain}
\affiliation{Laboratorio de Microscopias Avanzadas (LMA), Universidad de Zaragoza, E-50018, Zaragoza, Spain}

\author{Leyre Hern\'andez-L\'opez}
\affiliation{Instituto de Nanociencia y Materiales de  Arag\'on (INMA), CSIC-Universidad de Zaragoza, Zaragoza 50009, Spain}
\affiliation {Departamento de F\'{\i}sica de la Materia Condensada, Universidad de Zaragoza, E-50009 Zaragoza, Spain}

\author{Mikhail M. Otrokov}
\affiliation{Centro de F\'{\i}sica de Materiales CSIC/UPV-EHU-Materials Physics Center, Manuel Lardizabal 5, E-20018 San Sebasti\'an, Spain}
\affiliation{Donostia International Physics Center, Paseo Manuel de Lardizabal 4, E-20018 San Sebastian, Spain}
\affiliation{IKERBASQUE, Basque Foundation for Science, E-48011 Bilbao, Spain}

\author{Ignacio Piquero-Zulaica}
\affiliation{Centro de F\'{\i}sica de Materiales CSIC/UPV-EHU-Materials Physics Center, Manuel Lardizabal 5, E-20018 San Sebasti\'an, Spain}
\affiliation{Physics Department E20, Technical University of Munich, 85748 Garching, Germany}

\author{Adriana Candia}
\affiliation{Instituto de Nanociencia y Materiales de  Arag\'on (INMA), CSIC-Universidad de Zaragoza, Zaragoza 50009, Spain}
\affiliation{Instituto de Desarrollo Tecnol\'ogico para la Industria Qu\'{\i}mica (INTEC-UNL-CONICET),3000, Santa Fe, Argentina}
\affiliation{Instituto de  F\'{\i}sica del Litoral, Universidad Nacional del Litoral (IFIS-UNL-CONICET), 3000, Santa Fe, Argentina}

\author{Pierluigi Gargiani}
\affiliation{ALBA Synchrotron Light Source, E-08290 Cerdanyola del Vallès, Spain}

\author{David Serrate} 
\affiliation{Instituto de Nanociencia y Materiales de  Arag\'on (INMA), CSIC-Universidad de Zaragoza, Zaragoza 50009, Spain}
\affiliation {Departamento de F\'{\i}sica de la Materia Condensada, Universidad de Zaragoza, E-50009 Zaragoza, Spain}
\affiliation{Laboratorio de Microscopias Avanzadas (LMA), Universidad de Zaragoza, E-50018, Zaragoza, Spain}

\author{Manuel Valvidares}
\affiliation{ALBA Synchrotron Light Source, E-08290 Cerdanyola del Vallès, Spain}

\author{Jorge Cerdà}
\affiliation{Instituto de Ciencia de Materiales de Madrid, CSIC, Cantoblanco, 28049 Madrid, Spain}

\author{Andrés Arnau}
\affiliation{Centro de F\'{\i}sica de Materiales CSIC/UPV-EHU-Materials Physics Center, Manuel Lardizabal 5, E-20018 San Sebasti\'an, Spain}
\affiliation{Donostia International Physics Center, Paseo Manuel de Lardizabal 4, E-20018 San Sebastian, Spain}
\affiliation{Departamento de Pol\'{\i}meros y Materiales Avanzados: F\'{\i}sica, Qu\'{\i}mica y Tecnolog\'{\i}a, Facultad de Qu\'{\i}mica UPV/EHU, 20080, Donostia-San Sebasti\'{a}n, Spain}

\author{Fernando Bartolom\'e} \email{fernando.bartolome@csic.es}
\affiliation{Instituto de Nanociencia y Materiales de  Arag\'on (INMA), CSIC-Universidad de Zaragoza, Zaragoza 50009, Spain}
\affiliation {Departamento de F\'{\i}sica de la Materia Condensada, Universidad de Zaragoza, E-50009 Zaragoza, Spain}

\begin{abstract}
Ferromagnetism is the collective alignment of atomic spins that retain a net magnetic moment below the Curie temperature, even in absence of external magnetic fields. Downscaling this fundamental property into strictly two-dimensions was proposed in metal-organic frameworks, but remained unachieved till present.
In this work, we demonstrate that extended, cooperative ferromagnetism is feasible in an atom thick two-dimensional metal-organic framework. This is remarkable since $\approx5\%$ of a monolayer of Fe atoms produces an out-of-plane easy-axis square-like hysteresis loop with large coercive fields over 2~Tesla and an extraordinary magnetic anysotropy. Such phenomena are driven by exchange interactions mainly through the molecular linkers, presenting a phase transition at $T_C\approx35$~K. Our findings settle a two decade search for ferromagnetism in two-dimensional  metal-organic frameworks and should  trigger the research in this and related fields.

\end{abstract}


\maketitle



\section*{Introduction}
Achieving extended two-dimensional (2D) ferromagnetism (FM) has been a long scientific yearn. The Mermin-Wagner theorem precludes this collective state  in its isotropic form when mediated by short-range exchange interactions at finite temperatures~\cite{Mermin1966}. Thus, this property was initially limited to bulk materials featuring dominant in-plane interactions~\cite{Yamada1972, Miedema1974} and ultrathin inorganic layers supported on metallic surfaces~\cite{Pietzsch2001, Pietzsch2004, Krause2007, Bickel2011}. Noteworthy was the discovery that random monodispersed rare-earth atoms directly adsorbed on MgO thin layers could retain a remanent spin due to the suppression of the spin-relaxation channels~\cite{Donati2016}. It is only very recently that the first evidences of `pure' (i.e. substrate decoupled), extended 2D-FM were obtained by exfoliation of van der Waals crystals~\cite{Gong2017, Huang2017}. In these cases the Mermin-Wagner limitations were surmounted by the significant presence of magnetic anisotropy within these materials. Such finding immediately sparked widespread attention due to the many envisioned fundamental and applied implications~\cite{Gong2019, Soriano2020, Wang2022}. Nevertheless, the initial frenzy  appeased since implementing layered van der Waals materials into devices turns out to be extremely challenging as the lateral size and exact thickness of these layers is difficult to control~\cite{Wang2022}.\\
\indent
Earlier candidates to exhibit 2D-FM were  metal-organic frameworks (MOFs) grown on metallic supports~\cite{Gambardella2009, Carbone2011}. 2D-MOFs  {\it a-priori}  contain all the essential ingredients to display 2D-FM:
Selectable metallic centers providing non-zero atomic spins and incomplete quenching of the magnetic orbital moment (because of the reduced point symmetry and chemical coordination)~\cite{Bartolome2010, Leedahl2019}, periodical spacing of these magnetic moments over the (non-magnetic) surfaces~\cite{Umbach2012}, tunable lateral separation between adatoms by the synthetically variable organic linkers~\cite{Schlickum2007}, reduced electronic overlap of these metal centers  with the substrate after 2D-MOF formation~\cite{Piquero2019}, and technically simple fabrication as they follow self-assembly protocols close to room temperature~\cite{Bartels2010, Dong2016}. Despite all these, 2D-MOFs have historically failed to explicitly exhibit  2D ferromagnetic remanence~\cite{Gambardella2009, Carbone2011,  Umbach2012,  Abdurakhmanova2013, Giovanelli2014, Arruda2020, Moreno2022, Umbach2014, Faraggi2015}.   Interestingly, many previous studies exhibited noticeable magnetic anisotropies, but neither spontaneous magnetization nor remanence were ever observed~\cite{Gambardella2009, Carbone2011,  Umbach2012,  Abdurakhmanova2013, Giovanelli2014, Arruda2020, Moreno2022, Umbach2014, Faraggi2015}. Moreover, the coupling among these metal centers is generally interpreted in terms of superexchange mechanisms through the organic ligands and less so by surface electrons~ \cite{Yosida1996}. \\
\indent
 In this work, we study the magnetism of an atomically thin 2D-MOF consisting of Fe atom centers and 9,10-dicyanoanthracene (DCA) molecular linkers forming a mixed honeycomb kagome lattice on Au(111). We take advantage of the monodomain and extended character of this network in the choice of the multitechnique approach we use, which consist of scanning tunnelling microscopy and spectroscopy (STM/STS), low energy electron diffraction (LEED), X-ray absorption spectroscopy (XAS) and X-ray magnetic circular dichroism (XMCD). We present indisputable evidence of long-range ferromagnetic order in a 2D-MOF with a Curie temperature ($T_C$) of $\approx 35$~K, which displays a very strong out-of-plane (OOP) magnetic anisotropy and a square hysteresis loop with a coercive field of {$\approx 2.1$}~T. Consistently, XMCD orbital sum rule yields a maximal unquenched OOP orbital magnetic moment for Fe centers of {$\langle m^l_z\rangle \approx 2$}~$\mu_B$. The magnetization as a function of temperature through the FM phase transition is well described by the honeycomb 2D Ising model involving strong uniaxial anisotropic magnetic centers. These results differ from those obtained for the inorganic system formed in absence of DCA molecules, which results in an array of Fe clusters on the Au(111) surface. We make use of first-principles density functional theory to explain the observation of FM at finite temperature in this 2D-MOF system, which exhibits a large single ion anisotropy at the Fe atoms and a significant exchange interaction across the molecular linkers with a minor contribution through the underlying substrate. \\


\section*{Results} 

\subsection*{Spatial and electronic structure of the 2D-MOF}
\begin{figure*}
\begin{center}
	 \includegraphics[width=1.0\textwidth,clip]{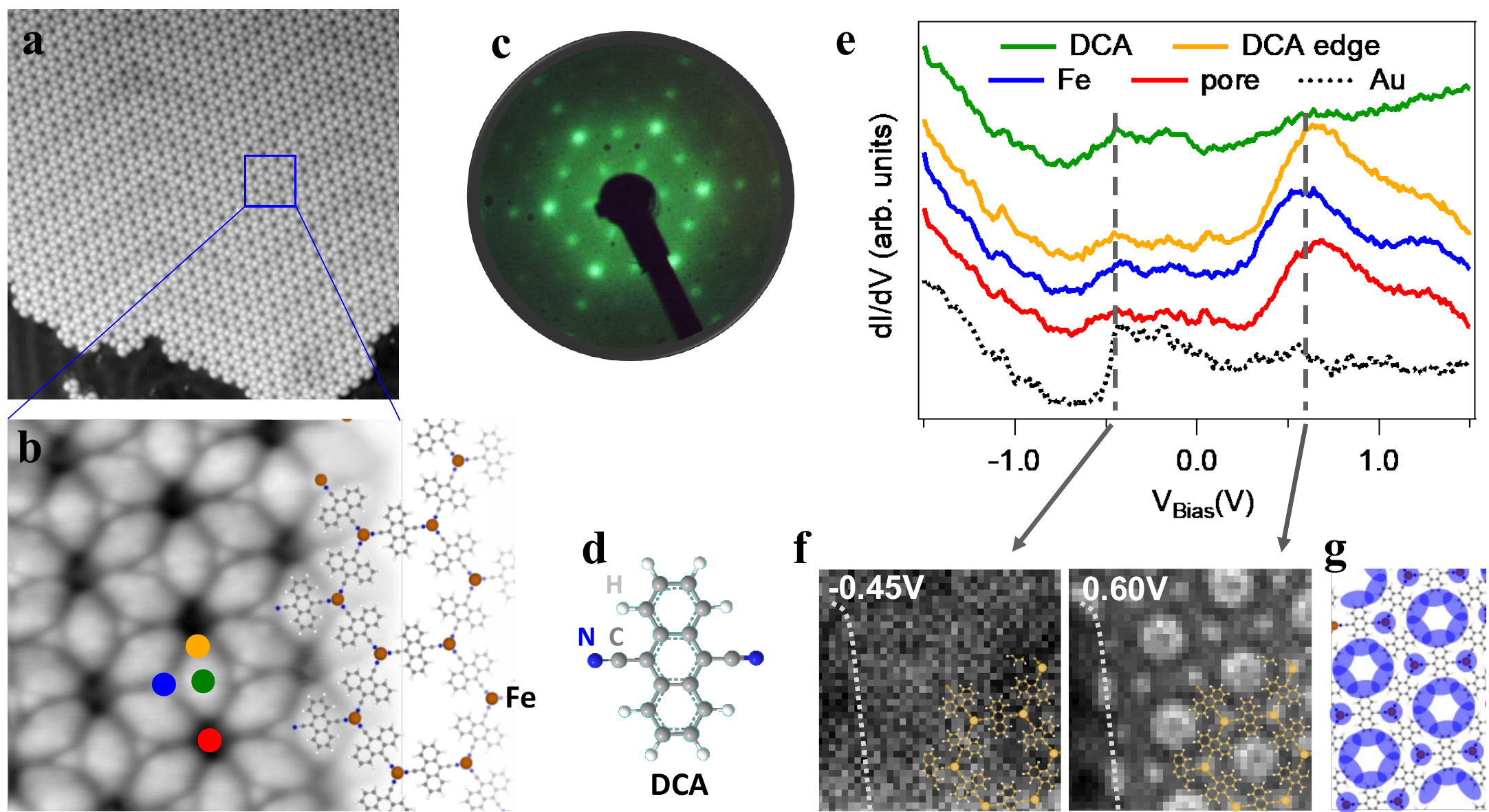}
 \end{center}
\vspace*{-8mm}
\caption[] {Atomic and electronic structure of the Fe+DCA/Au(111) network. 
	Overview (a) and close-up (b) images of the 2D-metal-organic lattice. A model of the MOF structure is overlaid in (b) with the molecules forming a kagome substructure and the Fe adatoms (orange-brown spheres) a honeycomb sublattice. 
(c) LEED pattern of this network at 19~eV and room temperature displaying a $(4\sqrt{3}\times4\sqrt{3})$R30$^{\circ}$ structure with respect to the underlying substrate. 
(d) Stick-ball model of 9,10-dicyanoanthracene (DCA) molecule (C in gray, N in blue and H in white). 
(e) STS averaged spectra extracted from a $dI/dV$ grid at the positions marked as circles of the same color in (b). A network (collective) electronic state is located around 0.60~eV. 
(f) $dI/dV$ isoenergetic maps at the energies indicated by the vertical discontinuous lines in (e). The gray dotted line on the right marks the edge of an island.
(g) Cartoon of the dominant spatial distribution (dark blue) of the network state identified at 0.60 eV.
STM parameters: (a) $50\times50$~nm$^2$, 100~pA; $-100$~mV; (b) $6\times6$~nm$^2$, 20~pA; $-1$~V; (e,f) Setpoint 100~pA, $-1$~V, $V_{rms} = 15.8$~mV, $f_{osc} = 817$~Hz). 
}
\label{figure1}
\end{figure*}

The self-assembled Fe+DCA lattice is formed by sequential deposition of DCA molecules and Fe atoms on Au(111). Prior to the metal evaporation, the molecules form compact islands (see {Fig. S1}) that evolve into the porous network under a stoichiometric Fe:DCA relation of 3:2 (see {Fig. S2} for the effects of deviation from this proportion). The network is perfected after a mild annealing at $373$~K for 10 minutes, generating large and monodomain network islands, similar to Cu+DCA/Cu(111)~\cite{Hernandez2021}.  Fig.~\ref{figure1}a shows a typical overview of this 2D-MOF, where the Fe centers form a honeycomb array and the DCA linkers a Kagome sublattice (cf. Figs.~\ref{figure1}b and S3e,f). This network does not destroy the herringbone reconstruction of the underlying Au(111), suggesting a weak surface-MOF interaction. The analysis of the LEED patterns (see Fig.~\ref{figure1}c) show a hexagonal network with $(4\sqrt{3}\times4\sqrt{3})$R30$^{\circ}$ structure with respect to the underlying substrate, resulting in unit vectors of 2~nm with first Fe neighbours distanced by $1.15$~nm.  \\ 
\indent
The electronic structure of this 2D-MOF is obtained by means of dI/dV grids from which STS spectra at selected network locations (Fig.~\ref{figure1}e) and dI/dV maps (Fig.~\ref{figure1}f) are extracted. These STS spectra clearly differ from the uncoordinated DCA (cf. {Fig. S1}). Two broad features call for our attention: The first, in the occupied region close to the Shockley state onset position ($\approx -0.50$~V), and the second at the unoccupied region centered at  $0.60$~V. In the dI/dV map at $-0.45$~V this first state is rather featureless both on the metal and throughout the network, revealing a substrate origin. This is supported by the fact that no distinct confined state could be detected at the MOF's pore centers, which further supports the  weak surface-network interaction. 
Contrarily, the dI/dV map at $0.60$~V shows distinct features throughout the network, primarily  at the DCA edges and at the metal centers, as sketched in  Fig.~\ref{figure1}g.
This has been identified as the fingerprint of an extended network multi-band in the related  Cu+DCA/Cu(111) system~\cite{Hernandez2021}. Such electronic hybridization has also been reported in the Cu+BHT network, revealing  superconducting properties~\cite{Takenaka2021}. \\

\subsection*{Magnetic characterization}
\begin{figure*}
\begin{center}
	 \includegraphics[width=1.0\textwidth,clip]{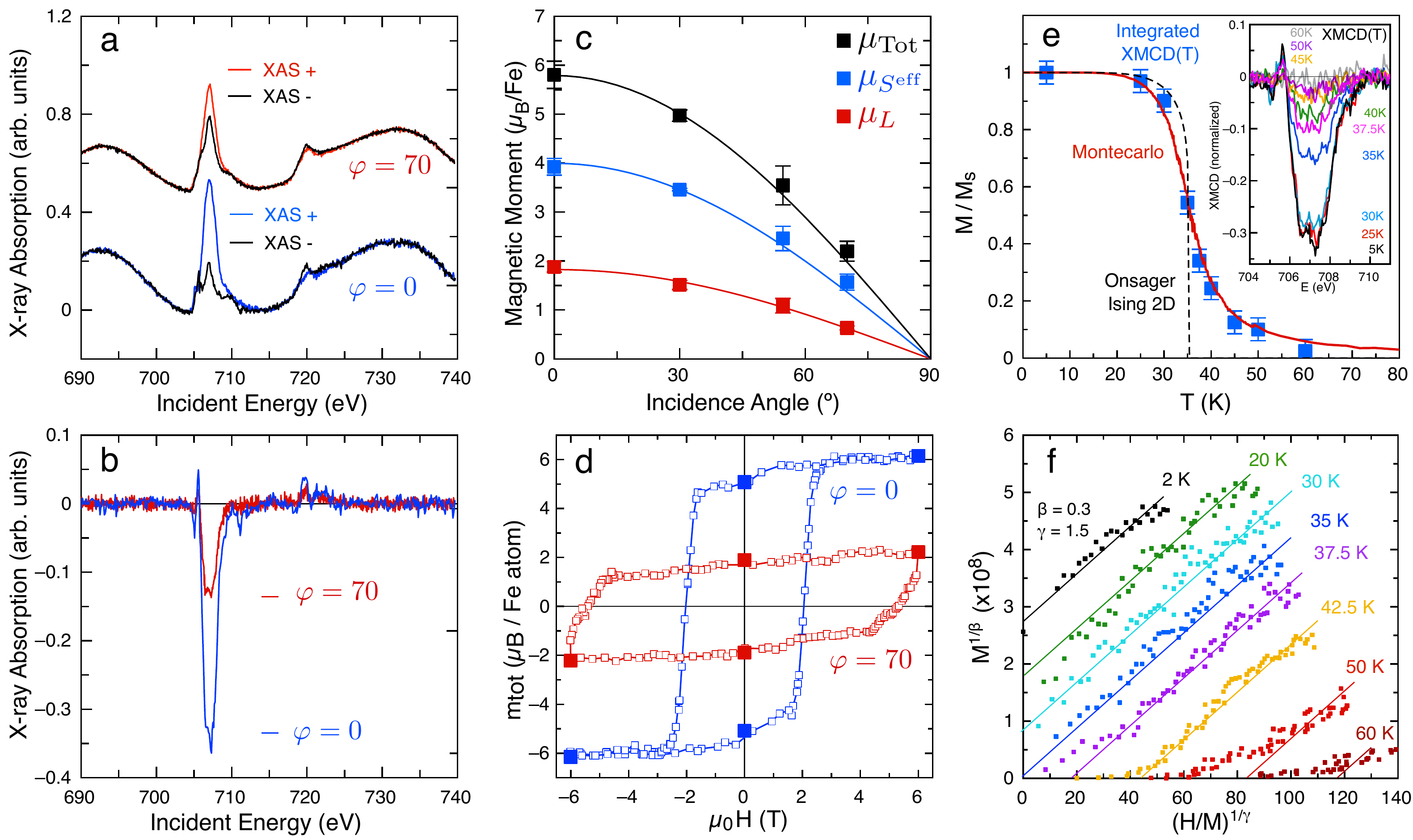}
 \end{center}
\vspace*{-3mm}
\caption[] 
{
Experimental demonstration of 2D-Ferromagnetism of the Fe+DCA network on Au(111). 
(a) XAS (displaced vertically for clarity) and corresponding (b) XMCD spectra acquired with circularly right ($I^+$) and left ($I^-$) polarized X-ray light for normal (0$^{\circ}$) and grazing (70$^{\circ}$) incidence at the $L_{2,3}$ edges of Fe.  The Fe L$_{2,3}$ XAS sits on top of the Au EXAFS background. A strong out-of-plane magnetic anisotropy is evidenced by inspection. 
(c) Angular dependence of the orbital (red), effective spin (blue) and total (black) magnetic moments obtained from the sum rules. The lines follow a cosine relation: $\mu_{\kappa}^{\varphi=0} \cos(\varphi)$ with $\kappa = L$, $S_{\mathrm{eff}}$, and Total.
(d) Hysteresis loops (open symbols) obtained at the $L_3$ edge of Fe at normal ($\varphi=0^{\circ}$, blue) and grazing ($\varphi=70^{\circ}$, red) incidence. The solid symbols are the result of applying the sum rules to the  XAS and XMCD spectra obtained after conveniently cycling the field from $6$~T to zero to  $-6$~T and back for both incidence angles. 
(e) Integrated area (normalized to the saturation value) of the $L_3$ XMCD main peak measured in normal incidence under low fields ($H=0.1$~T) as a function of temperature from $T=5$~K up to 60 K. The original XMCD data are shown in the inset. The Onsager 2D Ising analytical solution (dashed line) and a Monte Carlo simulation, both for a honeycomb lattice are shown, satisfactorily describing the experimental data for $J \approx  2$~ meV. The experiment was performed crossing $T_C$ heating up and then cooling down, exhibiting reversibility. Indeed, such $T_C$ reversibility was reproduced on a second Au(111) substrate under $H=0.05$~T (cf. Fig.~S6).   
(f) Modified Arrott-Noakes plot of isotherms with $\beta=0.30$ and $\gamma = 1.51$ corresponding to a long-range 2D-Ising model. \cite{Tateiwa2018}. 
}
\label{figure2}
\end{figure*}

Our results evidence perfect crystalline quality  and collective electronic states in this 2D-MOF.   Therefore, its magnetic properties can be unveiled using spatially averaging synchrotron-based techniques. A true magnetic signal probed by  XAS and XMCD requires that we prevent Fe cluster formation (cf. {Fig. S2} and {Fig. S3}). Thus, we target untraceable Fe undercoverage samples exclusively leading to single metal centers surrounded by three N atoms (from different cyano groups) which results in a local three-fold symmetric arrangement ($C_{3v}$). This is a single layer MOF system, so interlayer coupling is discarded due to the non-magnetic character of the underlying Au substrate. Fig.~\ref{figure2}a and b show XAS and XMCD spectra respectively, acquired at the L$_{2,3}$ edges of Fe at normal ($\varphi=0^{\circ}$) and grazing ($\varphi=70^{\circ}$) incidence {(see SI for experimental details)}. Both XAS and XMCD show narrow peak contributions, which are reminiscent of systems featuring monodispersed Fe atoms on surfaces \cite{Gambardella2002a, Pacchioni2015}, or embedded in other 2D-MOFs~\cite{Umbach2012}, or forming part of molecules \cite{Bartolome2010}. Indeed, the spectra clearly suggest a Fe(II) oxidation state (see for example Ref.~\cite{Kowalska2017}). To further confirm the absence of Fe cluster formation in the 2D-MOF, we directly deposit the same amount of Fe on the clean Au(111) substrate (without DCA molecules) and measure Fe L$_{2,3}$ XAS and XMCD under identical experimental conditions. The Fe/Au(111) results are shown in {Fig. S4}. Distinctly different XAS and XMCD spectra are obtained compared to the ones of Fig.~\ref{figure2}a which exhibit the typical smoother and broader metallic Fe L$_{2,3}$ lineshapes~\cite{Boeglin2002, Pacchioni2015}. These spectra evidence that Fe adatoms are assembled into small clusters nucleating at the herringbone elbows (see Fig. S2a)~\cite{Delga2011, Ohresser2001}. A direct inspection of Fig.~\ref{figure2}b reveals a strong OOP anisotropy for our 2D-MOF. Identical results have been obtained in two different XAS/XMCD experimental runs, using different Au(111) crystals as substrate (cf. {Fig. S5 and S6}). Despite a common OOP character, the Fe clusters in the Fe/Au(111) sample show a considerably smaller anisotropy than that found on the MOF, and their L$_3$ vs L$_2$ branching ratio (relative peak intensities) is significantly lower than for the Fe+DCA network, evidencing a much smaller Fe orbital moment. Note that the Fe clusters are not fully comparable to the Fe centers in the 2D-MOF because their geometry and chemical environments are different. Particularly, the coordination differs not only laterally, but also affects its vertical interaction with the substrate.\\
\indent
We can quantitatively determine the orbital ($\mu_L^{z}$) and effective spin ($\mu^{\mathrm{eff}_z}_{S} = \mu_S - 7\mu_T^{z}$) magnetic moments for the 2D-MOF and the Fe clusters by using the X-ray magnetic dichroism sum rules~\cite{Thole1992,Carra1993} (described in the SI) with the XAS and XMCD spectra acquired at 4 different incidence angles ($\varphi=0^{\circ}, 30^{\circ}, 54.7^{\circ}$, and $70^{\circ}$).  The spectra were obtained in every case with the magnetic field parallel to the beam direction, such that XMCD yields the projection of the Fe magnetic moment along the direction of the applied field, $m(\varphi) = \vec{H} \cdot \vec{m}/H$. We use the nominal number of holes in the 3d band for Fe(II) ($n_h=4$) in these sum rules. As evidenced in Fig.~\ref{figure2}c, $\mu_L^{z}$ and $\mu^{\mathrm{eff}_z}_S$ are strongly  anisotropic, being much larger when the applied field is perpendicular to the 2D-MOF plane.  The obtained values in normal incidence are $\mu_L^{z}=1.88\pm0.02$~$\mu_{\mathrm{B}}$ and $\mu_S^{\mathrm{eff}_z}=4.02\pm0.04$~$\mu_{\mathrm{B}}$, consistent with a Fe(II) d$^6$ high-spin (HS) configuration with L = 2 and S = 2, carrying a large orbital moment $\langle m^l_z\rangle \approx 2\mu_{\mathrm{B}}$, which is responsible for the large magnetic anisotropy. Similar cases of d$^6$ HS strongly uniaxial angular momenta have been observed, such as single Fe(II) ions atop the nitrogen site of the Cu$_2$N lattice~\cite{Rejali2020} and Co(III) in one-dimensional cobaltate Ca$_3$Co$_2$O$_6$\cite{Leedahl2019}.\\
\indent
The four incidence angles graphed in Fig.~\ref{figure2}c were selected to allow direct separation of the isotropic spin moment, $\mu_{S}$, from the quadrupolar term $- 7\mu_T^{z}$. In particular, if the magnetic moments rotate with the field as the angle-dependent experiment is performed, at the so-called  `magic angle' $\varphi = 54.7^{\circ}$, the intra-atomic dipolar contribution $\mu_T^{z}$ cancels, allowing a direct measure of $m_S$ \cite{Stohr1995,Bartolome2010}. However, this is not the case in Fe+DCA/Au(111) since we find that both $\mu_L^{z}(\varphi)$ and $\mu^{\mathrm{eff}_z}_S(\varphi)$ vary as $\cos(\varphi)$. Such behaviour can only be rationalized if the magnetic moments stay perpendicular to the 2D-MOF surface even for the extreme case of $\varphi =70^{\circ}$ and applied fields of 6T (see Fig.~S7), thereby evidencing a robust Ising  character with OOP quantization axis. \\
\indent
Such marked OOP robustness should stand out when recording hysteresis loops. Therefore, we acquire the magnetization curves at $T=3$~K by measuring the XMCD intensity at the fixed photon energy of the L$_3$-edge  as a function of the applied magnetic field for normal ($\varphi=0^{\circ}$) and grazing ($\varphi=70^{\circ}$) incidence (see Fig.~\ref{figure2}d) . Remarkably, the 2D-MOF presents a square-like open hysteresis loop with a huge coercive field value ($\approx 2.1$~T out-of-plane and $\approx 5.4$~T at $\varphi = 70^{\circ}$) and a remanence slightly above $80\%$ of the saturation value. Full zero field XMCD spectra at $\varphi=0^{\circ}$, and $70^{\circ}$ were measured directly after saturation with $\mu_0 H = \pm 6$~T, with sum-rules yielding the values shown as solid symbols in Fig.~\ref{figure2}d, consistently scaling with the hysteresis loop curve displayed. It is worth noting that the $\varphi=0^{\circ}$ hysteresis loop can be calculated from the $\varphi=70^{\circ}$ one (and viceversa), by simply scaling the applied field as $H_{0^{\circ}} = H_{70^{\circ}}\cos(70^{\circ})$ and the magnetization as $M_{0^{\circ}} = M_{70^{\circ}}/\cos(70^{\circ})$ (see Fig.~S7). In other words, the OOP component on the hysteresis loops are identical for all measured angles when projecting in that direction  the applied magnetic field, evidencing once more  the intense uniaxial character of the magnetic moments of Fe(II) in this 2D-MOF.\\
\indent
To  determine whether the open hysteresis loop reflects a slow relaxation single-atom process~\cite{Donati2016} or a truly 2D ferromagnetic cooperative phase transition, we  measure low field ($\mu_0 H=0.1$~T) XMCD curves as a function of temperature. The spectra at the L$_3$ edge are shown in the inset of Fig.~\ref{figure2}e, where we observe a sudden reduction of the XMCD signal occurring around $T_C\approx 35$~K. This critical temperature is evident in  Fig.~\ref{figure2}e when plotting the absolute value of the integrated area of the XMCD main peak at the L$_3$ edge (from $705$ to $712.5$ eV). The 2D Onsager's analytical solution (at zero field)~\cite{Onsager1944} scaled for $T_C=35$~K is shown for comparison,  yielding an exchange interaction of $J=[T_C\cdot k_B\cdot \log(\sqrt{3}+2)]/2\approx+1.98$~meV. A Monte Carlo simulation of the spontaneous magnetization under this low field ($\mu_0 H=0.1$~T) with that particular exchange interaction constant on a simple honeycomb network of Ising spins with nearest neighbour interactions  quite satisfactorily fits our experimental data. Note that this exchange constant is  considerable when compared with those found in other 2D-MOFs ($J\leq0.27$~meV)~\cite{Umbach2012, Giovanelli2014, Abdurakhmanova2013, Blanco2018}.
\\
\indent
To elucidate the type of ferromagnetic phase transition and the nature of spin-spin interactions existing in our 2D-MOF we extract the critical exponents from the measured XMCD magnetization curves at different temperatures. We expect these exponents to approximate to the `canonical' 2D Ising model ($\beta = 1/8 = 0.125$ and $\gamma = 7/4 = 1.75$) or to  well-established cases of strongly uniaxial 2D system with long-range interactions, such as URhAl ($\beta = 0.287$ and $\gamma = 1.47$)~\cite{Tateiwa2018}. Indeed, the modified Arrot-Noakes plot~\cite{Bedoya2021} displayed in Fig.~\ref{figure2}f provides values within range ($\beta = 0.3$ and $\gamma = 1.5$) that validate this 2D-MOF as a $d=2$ Ising system.  Note that the XMCD spectra remained proportional to the magnetisation since their lineshape did not change during the magnetic transition.  Considering the low amount of Fe centers present in this system ($\approx5\%$ of a monolayer) the experimental determination of these critical values is a remarkable achievement  in itself~\cite{Barla2016}. 
Furthermore, the approximate temperature independent behaviour regarding the slope of the Arrot-Noakes curves at higher fields (cf. Fig.~\ref{figure2}f) neatly highlights the cooperative character of the FM below $T_C \approx 35$~K. In short, all our XMCD datasets presents this 2D-MOF as a archetype example of a two-dimensional ferromagnet.\\
\indent
At this point, it is worth noting that Fe magnetic moments are strongly  dissimilar when comparing Fe+DCA/Au(111) and Fe/Au(111) samples, even if they share OOP anisotropy (cf. Fig.~\ref{figure2}b and Fig.~S4a and Table S1). Remarkably, the hysteresis loops measured on the Fe clusters (see Fig.~S4d) are rather similar at $\varphi=0^{\circ}$ and $70^{\circ}$ and do not reach saturation at the highest accessible external field ($6~$T). Moreover, the butterfly-shape hysteresis found in grazing incidence (closure at $0~$T) of Fe/Au(111) is characteristic of paramagnetic (or superparamagnetic) systems with long relaxation times \cite{Jiang2011, Zhu2011}. \\

\indent
\subsection*{Theory and discussion}
\begin{figure*}
\begin{center}
	 \includegraphics[width=0.9\textwidth,clip]{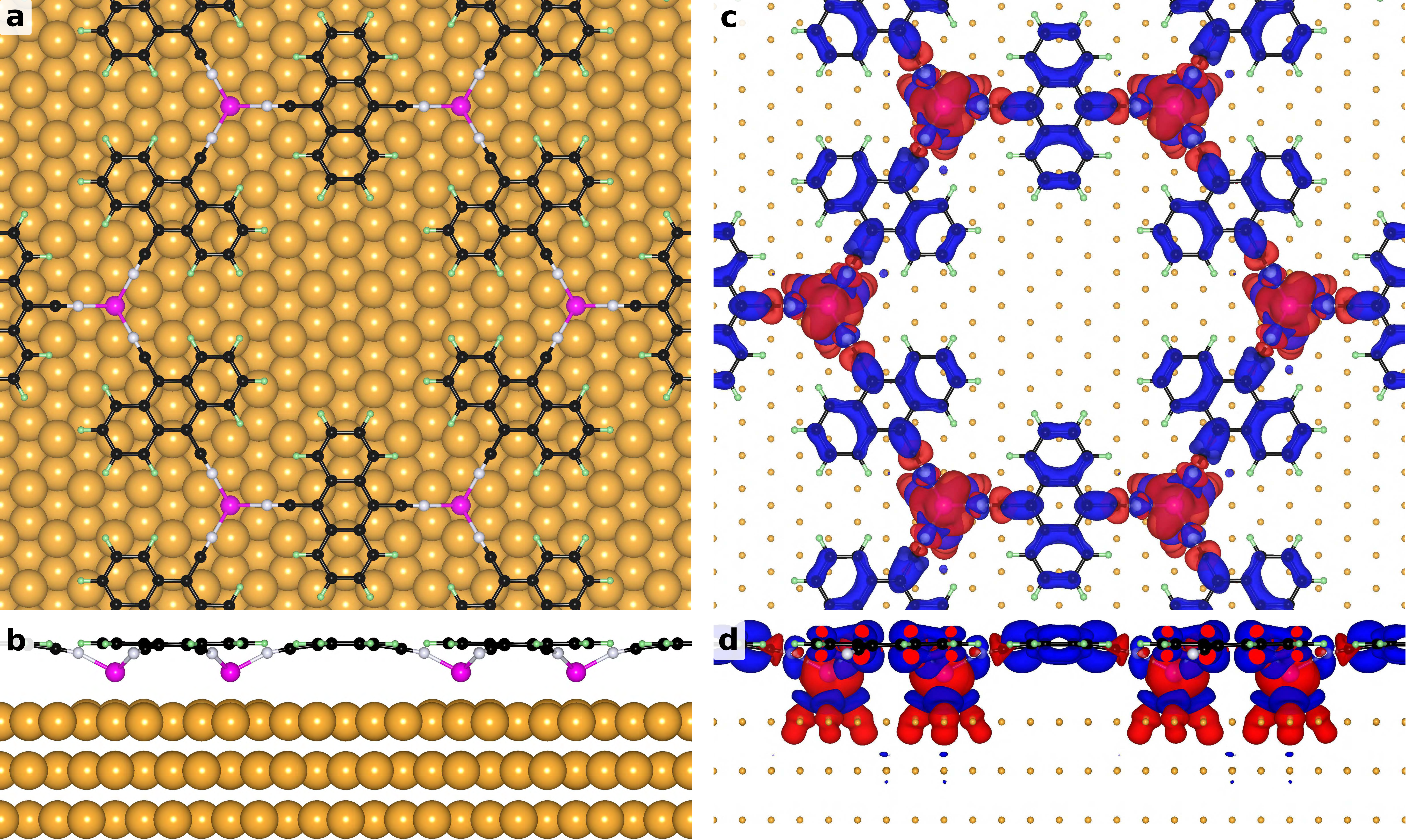}
 \end{center}
\vspace*{-3mm}
\caption[] {DFT calculations of the Fe+DCA/Au(111) system.
Top (a) and side (b) views of the 2D-MOF optimized structure with Au, Fe, C, N and H atoms in gold, purple, black, white, and green colors, respectively. The corresponding spin density isocontours [red (blue) denotes the majority (minority) spin component] are shown in panels (c) and (d). This spin density reveals a predominant coupling between Fe magnetic moments through the DCA ligands and a sizable spin polarization of the Au atoms below the Fe atoms. Note that the absence of significant spin density connecting the Au atoms evidences a limited coupling mediated by the metal surface.
}
\label{figure3}
\end{figure*}

 To shed light on these results we perform first-principles density functional theory calculations (details in SI). We start by structurally optimizing the Fe+DCA/Au(111) system and obtain a non-planar MOF geometry with the Fe atom and DCA backbone planes above the Au surface at $\approx2.5$~\AA\,  and $\approx3.9$~\AA\, respectively (cf. Fig. \ref{figure3}a,b). Independently of this buckled geometry, the MOF retains its two-dimensional character with respect to the magnetic properties of interest since the Fe atoms are coplanar, very much like the transition metal atoms in other 2D-systems, e.g., CrI$_3$. The vertical distortion affects the cyano groups that bend downwards to the Fe centers. Importantly, the shortest Fe-Au distance is 2.86 \AA\, that is 4\% larger compared to the Au$_{0.5}$Fe$_{0.5}$ bulk alloy~\cite{Raub1950}, which reduces the Fe-Au hybridization, as anticipated experimentally. Such  hybridization reduction related to the separation of the coordination atom  from the substrate is common to other 2D-MOFs~\cite{Piquero2019, Piquero-Zulaica2022}.    \\
 \indent 
Total-energy calculations of this Fe+DCA/Au(111) structure result in a ferromagnetic isotropic exchange coupling constant of $J \simeq$ 0.7 -- 1.3~meV  (Hubbard $U_{eff}$ parameter dependent, see Table S2) and a positive easy-axis magnetic anisotropy energy of $E_a\simeq0.6$~meV per Fe atom (see SI), contributions from the single ion anisotropy $D$ and anisotropic exchange interaction $\lambda$ being about $2DS^2 \simeq$ 1.5 meV and $3\lambda S^2 \simeq -0.25$ meV (i.e., $E_a \simeq  DS^2+3\lambda S^2/2$), respectively. This leads to FM with an OOP orientation of the local magnetic moments of {$3.7 \mu_B$} ($S=2$ state), which  nicely matches our experimental results  (compare tables S1 and S2). Note that ferromagnetic coupling seems to be rather general in metallic MOFs~\cite{Gambardella2009, Carbone2011,  Umbach2012,  Abdurakhmanova2013, Giovanelli2014, Arruda2020, Moreno2022, Umbach2014, Faraggi2015}. However,  in contrast to what is here reported, no open hysteresis loops were  detected in any of the previously studied 2D-MOFs. \\
\indent
An intriguing question remaining is the definition of the  exchange channels that drive the system into such collective magnetic state. From the spin density isosurfaces (Fig.~\ref{figure3}c,d) it is evident that superexchange via DCAs  dominates, in agreement with previous works~\cite{Umbach2012, Bellini2011, Wegner2009}. 
However, we may expect some limited coupling mediated by the metal surface caused by the weak (but still non-negligible) Fe-Au hybridization that provides another exchange interaction channel between Fe localized spins~\cite{Wahl2007, Meier2008, Zhou2010, Girovsky2017}. Regrettably, such coupling mediated by the substrate cannot be separated from that through the ligands in the whole system, i.e. a full Fe+DCA/Au(111) calculation. Therefore, we perform first-principles calculations where we remove the DCA molecules from the optimized system, while keeping the honeycomb Fe array and the underlying Au(111) substrate fixed. In such a case, we find that the substrate contribution to the isotropic exchange is limited to only  $J_{substr} \simeq 0.03-0.07$ meV (ferromagnetic coupling). Note that this calculation without DCA molecules can only provide an order of magnitude estimate because the Fe atoms are in a different chemical environment to the Fe+DCA/Au(111) system, thereby introducing modifications to their magnetic state.
\\
\indent
For this 2D-FM to occur in a MOF, our first-principles calculations  identify two necessary conditions: (i) the existence of hybrid bands (see Fig. S8) with magnetic metal centers and organic linker orbital characters responsible for the ferromagnetic coupling between spin magnetic moments in the meV range, and  (ii) a large  magnetic anisotropy that translates into the opening of a gap in the spin wave excitation spectrum to overcome the Mermin-Wagner theorem~\cite{Mermin1966}. 
These two key ingredients  turn out to be truly remarkable in this Fe+DCA/Au(111) system, causing such finite temperature FM to emerge. Indeed, we find the same order of magnitude in $J$ and roughly equal magnetic anisotropy as in the CrI$_3$ monolayer ($J = 2.2$~meV and $E_a = 0.65$~meV~\cite{Lado2017}). Thus, it is rather unsurprising that our 2D-MOF also presents a similar $T_C$ as the one experimentally determined for the 2D-van der Waals ferromagnet ($T_C=45$~K~\cite{Huang2017}).\\

%
\section*{Conclusions}
\indent
In summary, we have studied the structural, electronic and magnetic properties of the Fe+DCA/Au(111) system. This 2D-MOF exhibits delocalized electronic states with limited interaction with the substrate that sets the stage for prevailing electronic overlap between molecules and metal centers and a robust ferromagnetic ground state.  Remarkably, we find open hysteresis cycles, an extraordinarily large uniaxial anisotropy, and 2D Ising critical behavior which translates into indisputable experimental evidence of a metal-organic 2D ferromagnet with a Curie temperature of T$_C\approx35~$K. The observed $\mu_L^z/\mu_S^{eff_z} \sim 1/2$ value, consistent with Fe(II) HS d$^6$ configuration (L=S=2) is large enough to lead to a high uniaxial magnetic anisotropy (easy OOP magnetization direction). Our first-principles calculations confirm both the order of magnitude of the FM exchange coupling and the sign of the magnetic anisotropy that are necessary to explain the observed magnetic order at a rather high temperature.  Importantly, the magnetic exchange constant found for this system is comparable to the highest reported for the ultrathin 2D-van der Waals ferromagnets~\cite{Gong2019, Soriano2020, Wang2022} and is certainly much higher than all other previous 2D-MOFs studied~\cite{Gambardella2009, Carbone2011,  Umbach2012,  Abdurakhmanova2013, Giovanelli2014, Arruda2020, Moreno2022, Umbach2014, Faraggi2015, Blanco2018}. 
Our findings settle over two-decades of search for 2D-FM in MOFs, thereby representing a clear advance in the transversal fields of magnetism and surface science. Similarly to the seminal work of isolating a single layer CrI$_3$~\cite{Huang2017}, we expect to boost the community's interest  in magnetic 2D-MOFs and trigger follow-up theoretical and experimental work capable of leading to new  ferromagnetic systems that exhibit even higher ordering temperatures with such ultra-low magnetic atom densities.\\

%
\section*{Acknowledgements}
With this work we pay homage to our college Jorge Cerdà that left us too early, may he rest in peace. We thank the BL-32 staff of ESRF for help in the initial experimental tests and Jesús Bartolomé for valuable help with Monte Carlo simulations. We further acknowledge the use of Servicio General de Apoyo a la Investigación-SAI of the Universidad de Zaragoza.  IPZ acknowledges support from Prof. Johannes V. Barth (TUM). We acknowledge financial support from Grant References No. PID2019-107338RB-C64, PID2019-103910GB-I00, PID2020-115159GB-I00, PID2020-116181RB-C32 and RED2018-102833-T funded by MCIN/AEI/10.13039/501100011033, by “ERDF A way of making Europe” and “European Union NextGenerationEU/PRTR”. Also financing from FlagEra SOgraphMEM PCI2019-111908-2 (AEI/FEDER) and the European Regional Development Fund (ERDF) under the program Interreg V-A Espa\~na-Francia-Andorra (Contract No. EFA 194/16 TNSI) is acknowledged.
We also thank the Basque Goverment Grant No. IT-1527-22 and the Aragonese Projects RASMIA E12\_20R and NANOMIDAS E13\_20R, co-funded by Fondo Social Europeo. The synchrotron radiation experiments were performed at BOREAS beamline under offical and inhouse proposals ID2020024265 and ID2021095444~\cite{Barla2016}.\\


\section*{Supplementary materials}
Experimental and Theoretical Methods\\
Brief description of the XMCD sum rules\\
The supplementary tables \\
The supplementary figures S1 to S8 referenced in the main text.\\
The supplementary references.\\

\section*{Author contributions}
J.L-C., L.H-L. and D.S. conducted the STM experiments and analysis; 
J.L-C., L.H-L., I.P-Z., A.C., P.G., M.V. and F.B. conducted the XAS / XMCD experiments, and L.H-L. and F.B. and analysed them; 
F.B. performed the Monte Carlo calculations;   
M.M.O, J.C. and A.A. performed the DFT calculations; 
J.L-C., F.B. and L.H-L. wrote the manuscript; 
J.L-C. and F.B. conceived the project; 
All authors contributed to the revision and final discussion of the manuscript.

\section*{Competing interests}
The authors declare no competing financial interests.


\pagebreak

\end{document}


\newcommand{\de }{$^{\circ}$}
\newcommand{\ADC}[1]{{\textcolor{orange} {\bf ADC: #1}}}
\newcommand{\JL}[1]{\textcolor{blue}{{\bf JL: #1}}}
\newcommand{\DSD}[1]{\textcolor{red}{{\bf DSD: #1}}}

\newcommand{\out}[1]{{\color{red}\sout{#1}}}
\newcommand{\add}[1]{{\color{blue}{#1}}}
\newcommand{\com}[1]{{\color{magenta} #1}}

\renewcommand{\thesection}{S\arabic{section}}
\renewcommand{\thetable}{S\arabic{table}}
\renewcommand{\thefigure}{S\arabic{figure}}
\renewcommand{\theequation}{S\arabic{equation}}
\newcommand{\FB}[1]{\textcolor{cyan}{{\bf FB: #1 }}}


\title{Supplementary Information:\protect\\ {Ferromagnetism on an atom-thick \& extended 2D-metal-organic framework}}

\author{Jorge Lobo-Checa} 

\author{Leyre Hern\'andez-L\'opez}

\author{Mikhail M. Otrokov}

\author{Ignacio Piquero-Zulaica}

\author{Adriana Candia}

\author{Pierluigi Gargiani}

\author{David Serrate} 

\author{Manuel Valvidares}

\author{Jorge Cerdà}

\author{Andrés Arnau}

\author{Fernando Bartolom\'e}

\maketitle
\setlength{\parskip}{12pt}
This document contains:
\begin{itemize}
	\item The Methods section with experimental and theory details.
	\item Brief description of the XMCD sum rules
	\item The supplementary tables 
	\item The supplementary figures referenced in the main text.
	\item The supplementary references.
\end{itemize}

\break

\section{Methods} 
{\bf Experimental details.}\\
- {\it Sample preparation.} 
 All the experiments were performed under ultra-high vacuum conditions. The base pressure in the preparation chamber was below $5\cdot10^{-10}$~mbar. Au(111) single crystals were prepared by repeated cycles of Ar$^+$ sputtering at $1.5$~kV and annealing at $750$~K. DCA molecules were evaporated from a Knudsen cell by thermal evaporation at a rate of $\approx0.03$~monolayer per minute (ML/min) at room temperature (RT). Fe atoms were subsequently evaporated from an e-beam evaporator at a rate of  $\approx0.005$~ML/min at RT. Finally an annealing up to $370$~K was performed in all samples to improve the MOF long-range order.\\
 \indent
The full network ML coverage of DCA was calibrated by directly depositing the molecules on a Cu(111) sample. On this surface the Cu+DCA MOF  spontaneously forms at RT and saturates the surface with the porous network~\cite{Hernandez2021}. Once the DCA evaporator was calibrated, the same deposition was done on Au(111) and the LEED was checked revealing the pattern shown in Fig. S1. To generate the Fe+DCA network we deposited  minute amounts of Fe atoms, annealed at $370$~K and checked the LEED at each evaporation step. We continued adding Fe until the LEED pattern eventually transformed into the pattern shown in Fig. S3. \\
\indent
Once the Fe and DCA evaporators were calibrated, we formed the DCA+Fe/Au(111) network on a single step. In the case of XAS and XMCD measurements, we took the precaution of reducing the Fe amount by $5\%$ to avoid Fe clusters forming on the surface (compare Figs. S2 and S3).\\

- {\it STM equipments.} 
 The initial topography and the determination of the electronic properties of the 2D-MOF were carried out at the Laboratorio de Microscopías Avanzadas (LMA) of the Universidad de Zaragoza using a low-temperature STM (LT-STM). This chamber has a base pressure better than $1\cdot10^{-10}$~mbar and operates at $\sim4$~K. A W tip was used in all cases. All the voltages are referred to the sample. All presented images were acquired with constant current mode.\\
 \indent
 A second setup equipped with a RHK STM head  was used at the BOREAS beamline in ALBA to judge the sample quality measured by XAS and XMCD. This equipment was operated at RT at a base pressure of  $3\cdot10^{-10}$~mbar. We acquired topographic images in constant current mode, as shown in Fig. S3. \\ 

- {\it XAS and XMCD setup.} 
 X-ray  absorption and dichroism experiments were performed at the Boreas beamline at ALBA synchrotron~\cite{Barla2016}. We did so in two different synchrotron runs with one year of time separation. The field  and temperature used for XAS and XMCD spectra collection were H=$6~$T and T$\sim3~$K, unless otherwise stated. We measured at the Fe $L_{2,3}$ edge fixing the magnetic field with the direction of the incident light.  The detection mode was total electron yield. The angle-dependent measurements were performed by rotating the sample about a vertical axis perpendicular to the synchrotron orbital plane, thereby varying the incidence angle $\varphi$ between the X-ray beam (and therefore the mangetic field) and the substrate normal. To minimize experimental artifacts  and reduce drift phenomena on the XMCD, we changed either the light helicity or the field direction during measurements. We increased the statistics by alternatively acquiring spectra for different helicities.\\
\indent 
It is important to indicate that beam damage existed. Indeed, we destroyed this network within seconds when we  illuminated the 2D-MOF with the full collimated beam. When that happened the XAS and XMCD spectra showed different lineshapes to Fig.~2a,b and, remarkably, the hysteresis loops closed. We learned to avoid this beam damage by reducing the full intensity by a factor of roughly 100 (detuning both the beam focus and also the undulators). Although the beam damage was avoided (several hours timescale), we further prevented adquisition errors by continuously  moving the spot on the sample surface (disc of $6.5$~mm diameter) to continuously access non-illuminated fresh sample spots.  \\
\indent
 
{\bf Theory details.}\\
Electronic structure calculations were carried out within 
density functional theory 
using the projector augmented-wave (PAW) method \cite{Blochl.prb1994} as implemented in 
the VASP code \cite{vasp1, vasp2}. The exchange-correlation energy was treated using 
the generalized gradient approximation \cite{Perdew.prl1996}. The Hamiltonian contained 
scalar relativistic corrections and the spin-orbit coupling was taken into account by 
the second variation method \cite{Koellingjpc1977}. The energy cutoff for the 
plane-wave expansion was set to 400 eV.  The Fe $3d$-states were treated employing the 
GGA$+U$ approach \cite{Anisimov1991} within the Dudarev scheme \cite{Dudarevprb1998}. 
The $U_\text{eff}=4, 5,$ and 6 eV values were used for our first-principles calculations. 
The atomic positions were fully optimized for each 
$U_\text{eff}$ and then the magnetic ordering, magnetic anisotropy, and electronic 
structure were studied (see Table \ref{table:jota}). It was found that neither the 
magnetic ground  state nor the electronic structure change qualitatively upon such a 
variation of  $U_\text{eff}$. 
Fe+DCA/Au(111) was simulated within the supercell approach using the experimentally found $(4\sqrt{3}\times4\sqrt{3})$R30\de{} periodicity;
the Fe atoms of the MOF were residing in the substrate's fcc hollow sites.
The cells contained a vacuum layer of a minimum of 10~\AA.

The free-standing Fe+DCA was fully optimized, i.e., lattice parameter and atomic coordinates corresponding to the minimal energy were found. For  all $U_\text{eff}$, the lattice parameter of Fe+DCA is roughly $a = 20.5$ \AA, which corresponds to Fe-Fe distance of 11.84 \AA. A compressive strain of about 2.55 \% is needed for its cell to match the $(4\sqrt{3}\times4\sqrt{3})$ Au(111) 
periodicity, as the  opitmized bulk lattice parameter of Au is 2.9153 \AA. This strain 
is accomodated via loss of the Fe+DCA planarity. Three Au 
layers were used to simulate the Au(111) substrate, so that the Fe+DCA/Au(111) cell
contained 224 atoms. The atoms of the lowermost Au layer were fixed during the 
structural relaxations, while the other two as well as the MOF were allowed to relax.
The atomic coordinates were relaxed using a force tolerance criterion for convergence 
of 0.01 eV/{\AA}. In all relaxations  the 2D Brillouin 
zone was sampled with a $2\times 2\times 1$ $\overline \Gamma$-centered $k$-mesh.
Using denser mesh eventually results into only $\sim$ 7 \% change of the exchange coupling constant,
while significantly slows down the calculation. All of the total-energy calculations where
performed using the $7\times 7\times 1$ $\overline \Gamma$-centered $k$-mesh.

We consider the 2D Heisenberg model with the out-of-plane easy-axis anisotropy.
\begin{equation}
\label{eq:h}
        H = -\frac{J}{2}\sum_{i,j}\vec{S}_i\vec{S}_j  - \frac{\lambda}{2}\sum_{i,j}S_i^z S_j^z - D\sum_{i}(S_i^z)^2,
\end{equation}
where $J$ is isotropic exchange coupling constant, $\lambda,D$ are the parameters of the exchange anisotropy and single-site anisotropy, respectively, while $i$ and $j$ denote nearest neighbors within the layer. While $J$ is  obtained via a scalar relativistic DFT calculation of ferromagnetic and antiferromagnetic configurations, determining $\lambda$ and $D$ requires a relativistic calculation of the latter configurations for both in-plane and out-of-plane moment directions. Structure-wise, we have used a periodic supercell with two Fe atoms, i.e., one pair of interacting spins, each Fe atom with its three nearest neighbours, two of them in neighbour unit cells.
This choice of supercell translates into an exchange coupling energy $JS^2$ per spin pair so we use  the expression $J=\Delta_{A/F}/6 S^2$ ($\Delta_{A/F} = E_{AFM} - E_{FM}$) to estimate the value of $J$. 

The $J$ values are presented in Table \ref{table:jota} for three different Hubbard $U_{eff}$ parameters. The coupling is ferromagnetic in all cases, its strength being sensitive to the $U_{eff}$ parameter. It is interesting to note that the isotropic exchange coupling between Fe atoms in the Fe+DCA/Au(111) is comparable to that between Cr atoms in CrI$_3$~\cite{Zhang.jmcc2015, Lado2017}, being the distance between Fe atoms roughly two times larger than between Cr atoms. The explanation is that the strength of the superexchange coupling, mediated by the I orbitals in CrI$_3$ or by the DCA molecular orbitals in Fe+DCA/Au(111), does not depend so strongly on the size of the orbital, but rather on the overlap between the DCA and Fe orbitals. Indeed, a similar behavior has been found in one dimensional metal-organic polymeric chains of Co and Cr atoms with organic QDI ligands, with exchange coupling constants of the order of meV~\cite{Wackerlin.acsn2022}.

The magnetic anisotropy energy, $E_a=E_{diff}+E_d$, was calculated taking into account 
the total energy differences, 
$E_{diff}=E_{in-plane} - E_{out-of-plane}$, and the energy of the classical 
dipole-dipole interaction, $E_d$. For the $E_{diff}$ calculation, a $k$ mesh of $7\times 7\times 1$ 
points was chosen. To calculate $E_{diff}$, the energies for three inequivalent 
magnetization directions [Cartesian $x$, $y$ (in-plane) and $z$ (out-of-plane)] were 
calculated and $E_{diff}$ was determined as the difference $E_{{in-plane}}$ -- $E_z$, 
where the $E_{{in-plane}}$ is the energy of the most energetically favorable in-plane 
direction of magnetization. The energy convergence criterion was set to 10$^{-7}$ eV 
providing a well-converged $E_{diff}$ (up to a few tenth of meV) while excluding 
"accidental" convergence. A cutoff radius of at least 20 microns was used to calculate 
dipole-dipole contribution $E_d$ to the magnetic anisotropy energy.
The obtained $E_a$ values are included in Table \ref{table:jota}. For all cases, the 
out-of-plane magnetization direction is favorable.  Using the previously described 2D Heisenberg model, we find the single ion anisotropy $D \approx 0.19$~meV and the anisotropic exchange $\lambda \approx -0.02$~meV, being both significantly smaller than the isotropic exchange with a higher contribution from single ion anisotropy.

We have also calculated the magnetic anisotropy of the free-standing monolayer 
in the absence of Au(111) substrate while freezing the buckled geometric structure and have 
found a $E_a$ value of the same sign and order of magnitude, although smaller (0.3 meV).

Some of the results calculated with VASP were verified against those obtained using the GREEN 
code~\cite{bib:Cerda97,bib:Rossen13} and its interface to the SIESTA DFT-pseudopotential 
package~\cite{bib:soler02} within the GGA$+U$ approach and a good agreement was found.

Finally, given the strong uniaxial single-ion anisotropy observed experimentally, we estimate the magnetization as a function of temperature using a basic honeycomb Ising Monte Carlo model essentially as implemented in Ref.~\onlinecite{Hasbun.2020}  
\begin{equation}
\label{eq:h}
	        H = -\frac{J^{{\mathrm{MC}}}}{2}\sum_{i,j}{S}_i{S}_j.
\end{equation}
Using $J^{{\mathrm{MC}}}=1.98$~meV and the experimental applied field (0.1T and 0.05T, respectively) the curves represented in Fig. 2(e) and Fig. S5(e) are obtained. Note that this $J^{{\mathrm{MC}}}=1.98$~meV exchange constant is in reasonable agreement with the one obtained by DFT (see Table~\ref{table:jota}).

\section{XMCD sum rules} 
 A XMCD spectrum is the difference between two XAS spectra obtained with opposite circular polarizations (whose helicity is typically parallel or antiparallel to the applied magnetic field). Typically, several spectra must be accumulated to obtain the desired statistics with acceptable signal-to-noise ratio. 

Once a high quality XMCD spectrum has been acquired, the orbital and spin contributions to the total magnetic moment can be, in principle, separately obtained. The tools allowing this are the magnetooptical sum rules derived by P. Carra, B.T. Thole and G. van der Laan in the 90's~\cite{Carra1993, Thole1992}. They can be expressed as:

\begin{equation}
    \mu_L=-\frac{2}{3} \frac{A+B}{C}n_h
        \label{eq:sumruleorb}
\end{equation}
\begin{equation}
    \mu^{\mathrm{eff}}_{S}=\frac{2B-A}{C}n_h
    \label{eq:sumrulespin}
\end{equation}

where $n_h$ is the number of holes in the final electron states band (the 3d band for the Fe L$_{2,3}$ absorption edges), $A$ and $B$ are the area enclosed under the $L_3$ and $L_2$ edges of the XMCD spectrum, respectively, and $C$ is the area enclosed in both edges of the XAS spectrum once the excitations to the continuum have been removed. A, B and C are calculated by integrating XAS and XMCD spectra. To remove the contribution from transitions to the continuum, typically a double-step function is subtracted from the XAS spectrum. With such a minute sample as the 2D-MOF we are dealing here, it is essential to treat consistently the whole series of data to obtain a robust result. 

The obtained parameter in equation \ref{eq:sumrulespin} is $\mu^{\mathrm{eff}}_{S}= \mu_S -\frac{7}{2}\mu_T$, where $\mu_S$ is the spin moment and $\mu_T$ the spin-quadrupole.
If the studied magnetic moment rotates with the magnetic field in an angle dependent experiment, the angle dependence of the orbital and effective spin magnetic moments is respectively given by ~\cite{Weller1995}: 
 \begin{equation}
    \label{eq:sumrule1}
        \mu_{L}(\varphi) = \mu^{z}_{L}\cos^2(\varphi) + \mu^{xy}_{L}\sin^2(\varphi)
 \end{equation}
 \begin{equation}
    \label{eq:sumrule2}
        \mu^{\mathrm{eff}}_{S}(\varphi) = \mu_{S}-7 [\mu^{z}_{T}\cos^2(\varphi) + \mu^{xy}_{T}\sin^2(\varphi)]
 \end{equation}

 where $\mu_S$ = $-2\langle S_z\rangle \mu_\mathrm{B}/\hbar$ is isotropic, but both  
 the orbital moment $\mu_{L}(\varphi)$ = $-\langle L_{z}^{\varphi}\rangle \mu_\mathrm{B}/\hbar$ and the dipole magnetic moment of the spin density distribution
 $\mu_{T}(\varphi) = \langle T_{z}^{\varphi}\rangle \mu_\mathrm{B}/\hbar$ are intrinsically anisotropic. Because the dipolar tensor is traceless, and therefore $\mu^{z}_{T} + 2 \mu^{xy}_{T}=0$, only four moment components in Eqs. \ref{eq:sumrule1} and \ref{eq:sumrule2} are independent.
A particular case worth mentioning occurs at the so-called `magic angle' of incidence ($\varphi^{m}$ = $54.7^{\circ}$), where $2\cos^2(\varphi^{m})$ = $\sin^2(\varphi^{m})$ so the dipolar term is canceled in  Eq.~\ref{eq:sumrule2}, yielding a direct measure of $\mu_{S}$~\cite{Stohr1995}.\\
\indent
	As an example of this procedure we have fitted the orbital, effective spin and total moment of the sample formed by Fe clusters on Au(111) to eqns. \ref{eq:sumrule1} and \ref{eq:sumrule2}, shown in Fig. S4b with the same color as the symbols. The `magic angle' measurement allows to determine $\mu_S = 2.85 \pm 0.15 \mu_\mathrm{B}$, and thus $\mu_T^z = 0.06 \pm 0.02 \mu_\mathrm{B}$, while $\mu_L^z = 0.60 \pm 0.05 \mu_\mathrm{B}$. However, the dependence of the Fe+DCA/Au(111) 2D-MOF is completely different exhibiting a $\cos(\varphi)$ dependence (dashed lines in Fig. S4b). Clearly the two systems obey different angular behavior due to the large differences on the anisotropy of the orbital moment which in the case of Fe+DCA/Au(111) anchors the spin to the easy OOP axis through spin-orbit interaction.
 

\newpage

\section{Supplementary tables}

%

\begin{table}[h]
	\begin{center}
		\begin{tabular}{|c|c|c|c|}
			\hline
			& $ \mu_L^{z}~(\mu_B)$ & $\mu^{\textrm{eff}}_S~(\mu_B)$ & $\mu_{tot}^{z}~(\mu_B)$ \\
		 
			\hline 
			Fe+DCA/Au(111) & $1.88\pm0.02$ & $4.02\pm0.04$ & $5.90\pm0.06$ \\
			Fe/Au(111) & $0.60\pm0.05$ & $ 2.85 \pm 0.15 $ & $3.85\pm0.20$ \\
			\hline
		\end{tabular}
	\end{center}
	\caption[Orbital and spin magnetic moments of Fe+DCA/Au(111)]{Orbital and spin magnetic moments of Fe in Fe+DCA/Au(111) and Fe/Au(111) obtained fitting the points in Figs.~2 and S4 and following Eqs. \ref{eq:sumrule1} and \ref{eq:sumrule2}.}
\label{tab:Fe_sumrules}
\end{table}

\hfill

\hfill

\begin{table}[h]	
	\centering
	\begin{tabular}{|c|c|c|c|c|c|}
		\hline

		 $U_{eff}$ (eV)      & $m$ ($\mu_B$) &  $S$  & $\Delta_{A/F}$ (meV per Fe pair) & $J$ (meV) & $E_a$ (meV per Fe atom) \\
        \hline                                                                                    
                   4        &     3.68      &   2   &      31.7                        & 1.321     & 0.55     \\
        \hline                                                                                          
                   5        &     3.75      &   2   &      25.9                        & 1.079     & 0.55     \\
        \hline                                                                                          
                   6        &     3.80      &   2   &      17.4                        & 0.725     & 0.66     \\
        \hline
	\end{tabular}
	\caption{ \label{table:jota}
	 Calculated magnetic parameters of Fe+DCA/Au(111) for three different Hubbard $U_{eff}$ values. 
	 $m$ $(\mu_B)$ -- local magnetic moment on the Fe atom,
	 $S$ -- spin, 
	 $\Delta_{A/F}=E_{AFM} - E_{FM}$ (meV per Fe pair), 
	 $J$ (meV) -- nearest neighbor Heisenberg exchange coupling parameter,
	 $E_a$ (meV per Fe atom) -- magnetic anisotropy energy.}
\end{table}

\clearpage 
\break

\section{Supplementary figures}


\newsavebox{\smlmat}
\savebox{\smlmat}{$\left(\begin{smallmatrix} 4.1 & 0 \\ 3.5 & 2.9\end{smallmatrix}\right)$}
\begin{figure}[h]
\centering
\includegraphics[width=0.7\linewidth]{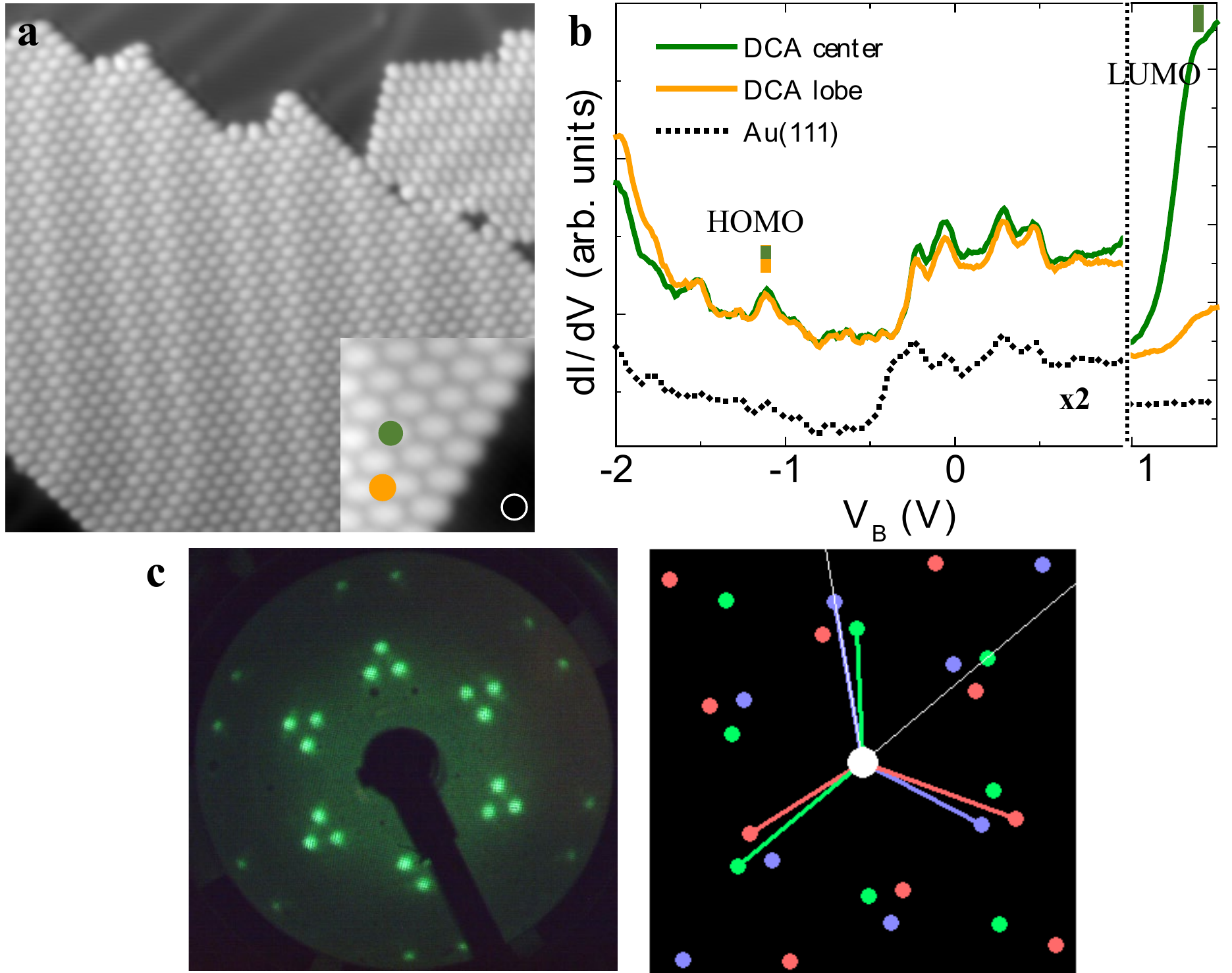}
\caption{Atomic and electronic characterization of DCA islands on Au(111).
a) STM overview showing that the molecules aggregate into compact islands without metal coordination. Note that the herringbone reconstruction is unaffected below the molecular self-assembly. 
b) STS acquired at the indicated positions of the inset image in a, with different intensity scaling above and below 1 V.
c) LEED characterization of the sub-monolayer DCA coverage on Au(111) without metal coordination and the corresponding simulated pattern. The arrangement corresponds to a rectangular oblique phase with vectors $a_1 = 1.18$~nm and $a_2 = 0.93$~nm and angle 50.8\de {} [matrix notation:~\usebox{\smlmat}] that contains three different domains. $E_{beam} = 19$~eV, $T_s$ = RT.
STM details: a) $30\times30$~nm$^2$, $I_t$ = 100~pA, $V_B = -1$~V. Inset: $5\times5$~nm$^2$, $I_t$ = 100~pA, $V_B = -1$~V. b) STS setpoint: $I_t = 100$~pA, $V_B = -1$~V; $V_{rms} = 9.6$~mV, $f_{osc} = 817$~Hz.
}
\label{figS1}
\end{figure}

\begin{figure}[h]
\centering
\includegraphics[width=1.0\linewidth]{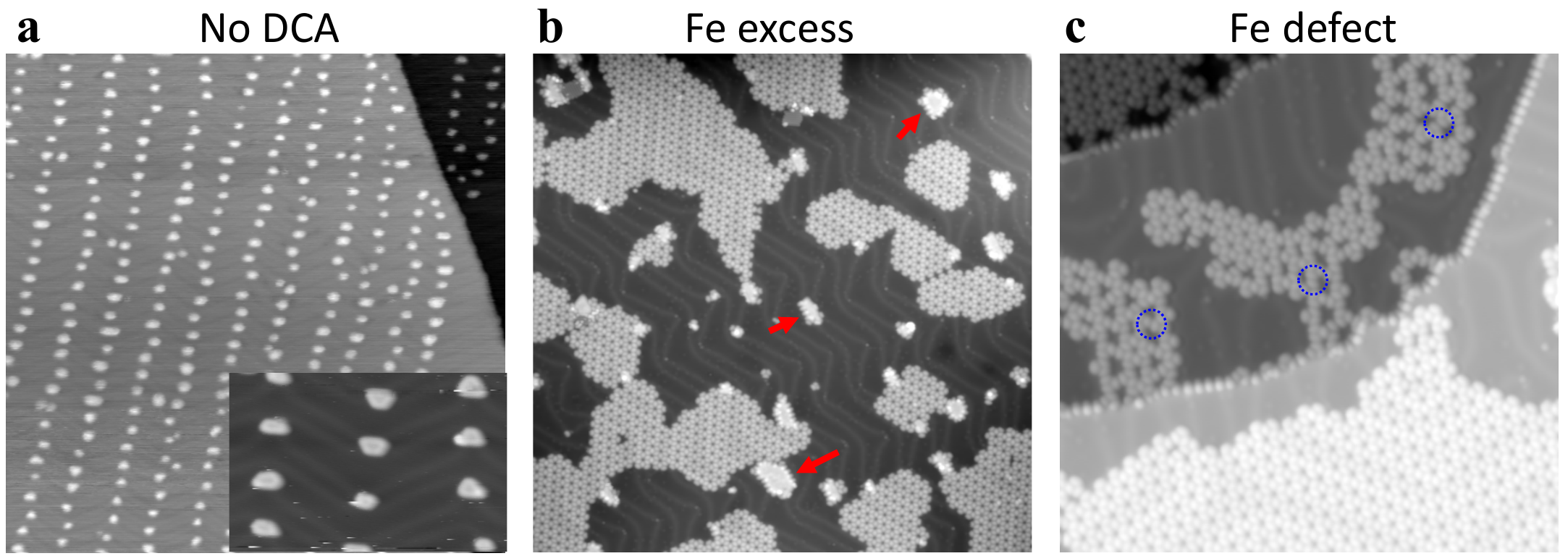}
\caption{Structural evolution modifying the Fe-DCA stoichiometry.
a) Direct deposition of Fe on Au(111) without DCA recorded at RT in the same stoichiometry as a full monolayer of the network. The atoms cluster into small triangular islands at the elbows of the herringbone reconstruction. 
b) Upon slight excess of Fe atoms compared to the DCA molecules, the porous network still dominantes yielding the highly regular structure with practically no internal defects. The Fe excess clusters into islands externally decorated by DCA as if they were step edges (indicated by red arrows). 
c) Lack of Fe on the surface results in irregular structures with enlarged pores originating from metal-uncoordinated cyanos. These “metal-free” cyanos are electrostatically bonded by dipole-dipole interactions to neighbouring DCAs (cf. blue dotted circles) so that the metal-organic trimers separate, thereby  enlarging laterally the pores.
STM details: a) $200\times200$~nm$^2$, $-0.003$~V, 100~pA, inset: $39\times25$~nm$^2$, 1.0~V, 10~pA; b) $100\times100$~nm$^2$, $-1.0$~V, 100~pA; c) $50\times50$~nm$^2$, $-1.0$~V, 100~pA.
}
\label{figS2}
\end{figure}

\begin{figure}[h]
\centering
\includegraphics[width=1.0\linewidth]{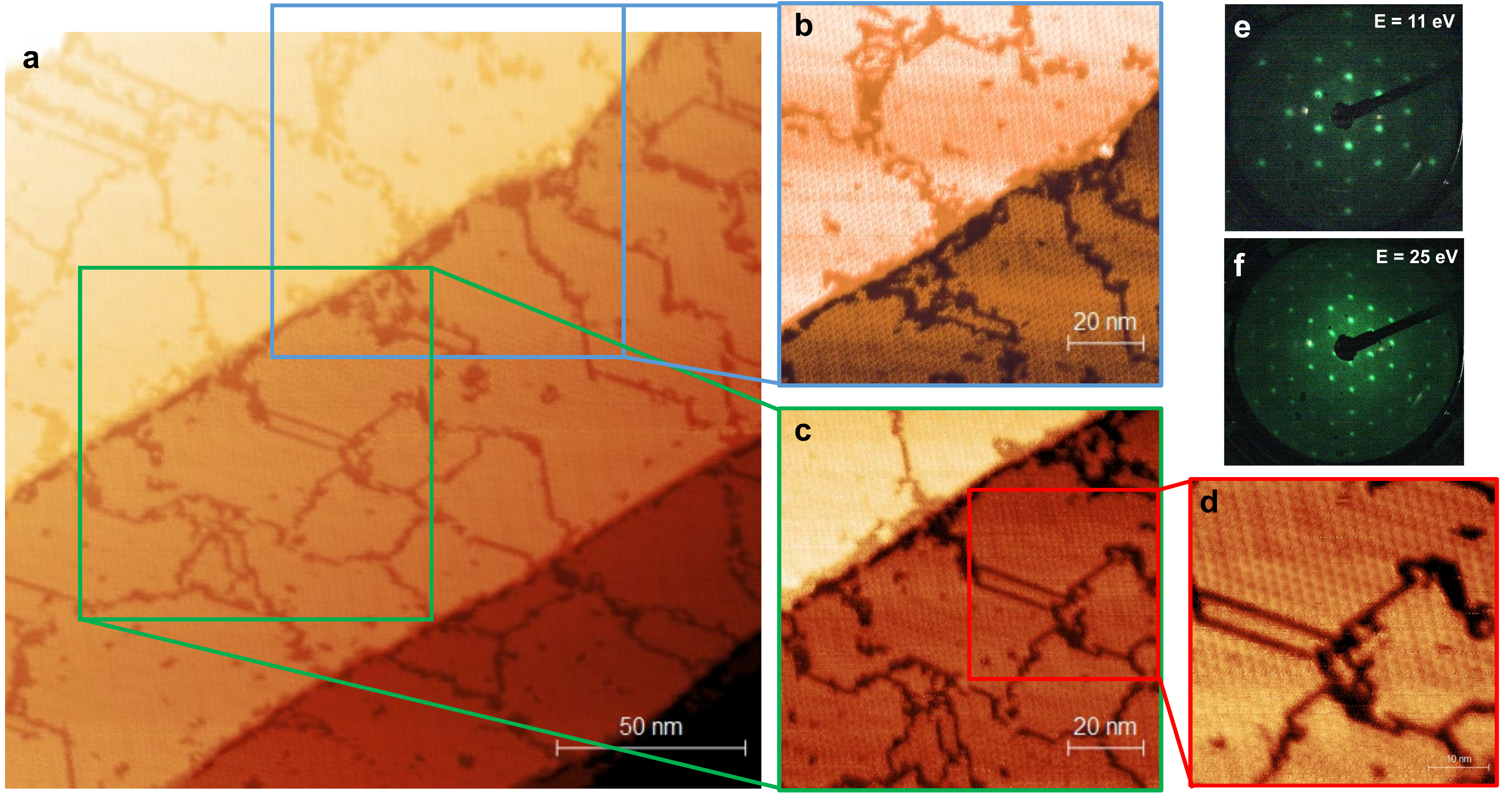}
\caption{Topography and LEED pattern of the sample measured by XMCD in Fig. 2 of the main manuscript acquired after the beam exposure. The STM overview a) and close ups b) - d) were acquired at room temperature after the XMCD acquisition. Note that no trace of Fe clusters can be detected on these STM images, exclusively showing the  Fe+DCA 2D-MOF. e) and f) display the room temperature LEED pattern at two different energies of the freshly prepared sample before being transferred to the XMCD setup. 
STM details: a) V= 0.8V, I = 400 pA, size $200\times200$nm$^2$; b) V= 0.8V, I = 400 pA, size $100\times100$nm$^2$; c) V= 0.5V, I = 400 pA, size $100\times100$nm$^2$; d) V= 0.5V, I = 400 pA, size $50\times50$nm$^2$.
}
\label{figS3}
\end{figure}


\begin{figure}
\begin{center}
	 \includegraphics[width=0.85\textwidth,clip]{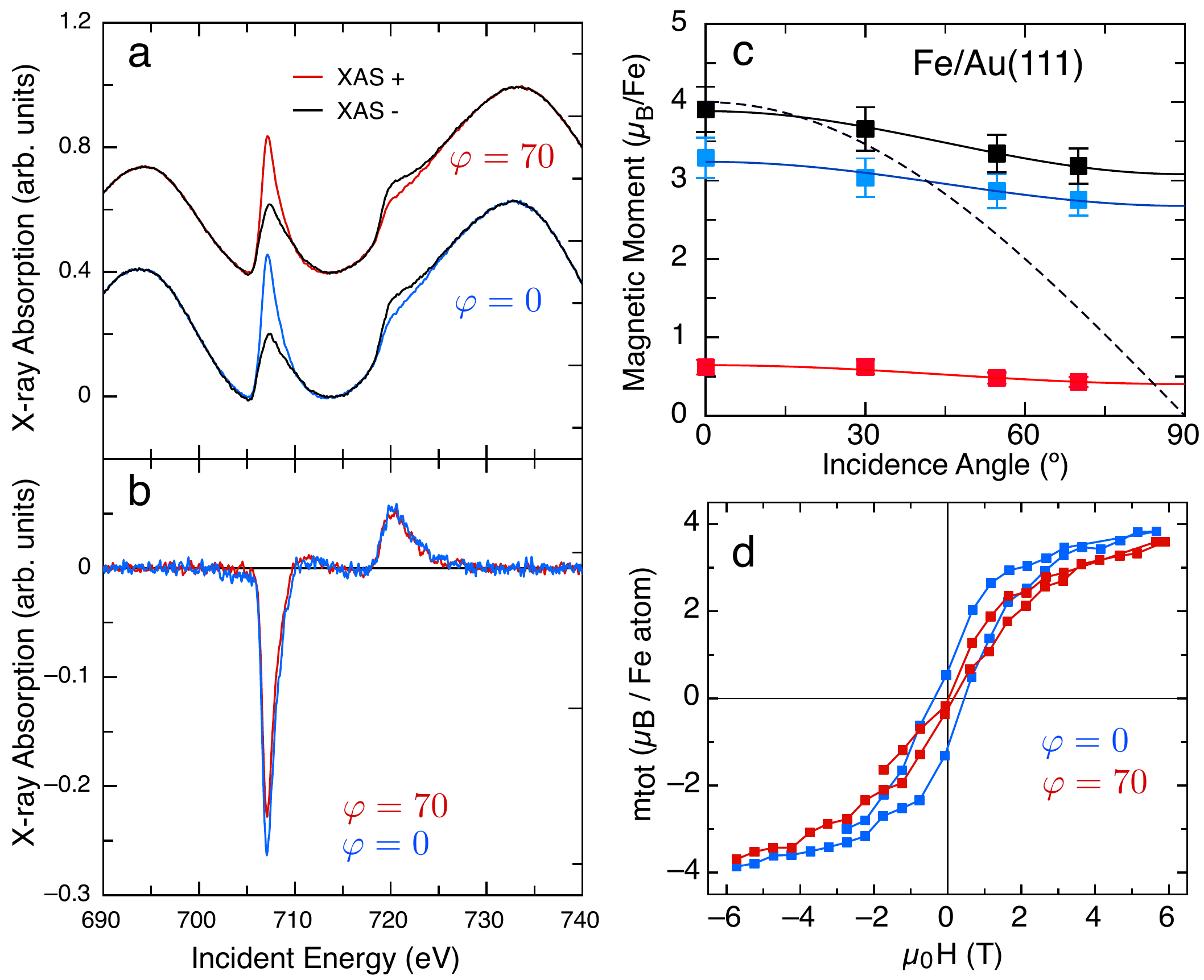}
 \end{center}
\vspace*{-3mm}
\caption[] {Magnetic experiments obtained for the Fe/Au(111) system at $\approx3$~K: (a) XAS and XMCD spectra obtained at normal (0\de) and grazing (70\de) incidence, (b) angular dependence of the orbital and spin effective moments, and (c) the hysteresis cycles at normal (0\de) and grazing (70\de) incidence. The amount of Fe on the surface was controlled to be identical to the experimental case of Fe+DCA/Au(111) network shown in Fig.~2 of the main manuscript. Several aspects must be underlined when comparing this graphs with the 2D-MOF system: First, XAS and XMCD lineshapes in (a) show a metallic character, very different to the ones shown in Fig. 2a. Second, the anisotropy of this Fe/Au(111), athough being OOP as in Fe+DCA/Au(111), is much weaker. Third, the orbital moment in the Fe/Au(111) (red squares in (b)) is practically angle independent and its magnitude is about 15\% of the total moment. This is quite high for metallic Fe, but is probably due to the high surface to volume ratio in the formed clusters. However, in the 2D-MOF its value is even higher (one third of the total magnetic moment) and clearly dominates the uniaxial character of the total moment (spin included). In contrast, in panel (b) the Fe clusters on Au(111) have slightly anisotropic magnetic moments (blue squares: effective spin, black ones: total moment).  Finally, the coercive field of Fe on Au(111) is significantly lower than in the 2D-MOF: it is just 0.5~T for normal incidence, and the grazing incidence curve displays a butterfly hysteresis loop that is probably related with slow relaxation of superparamagnetic units.
}
\label{FigS5}
\end{figure}

\begin{figure}
\begin{center}
	 \includegraphics[width=1.0\textwidth,clip]{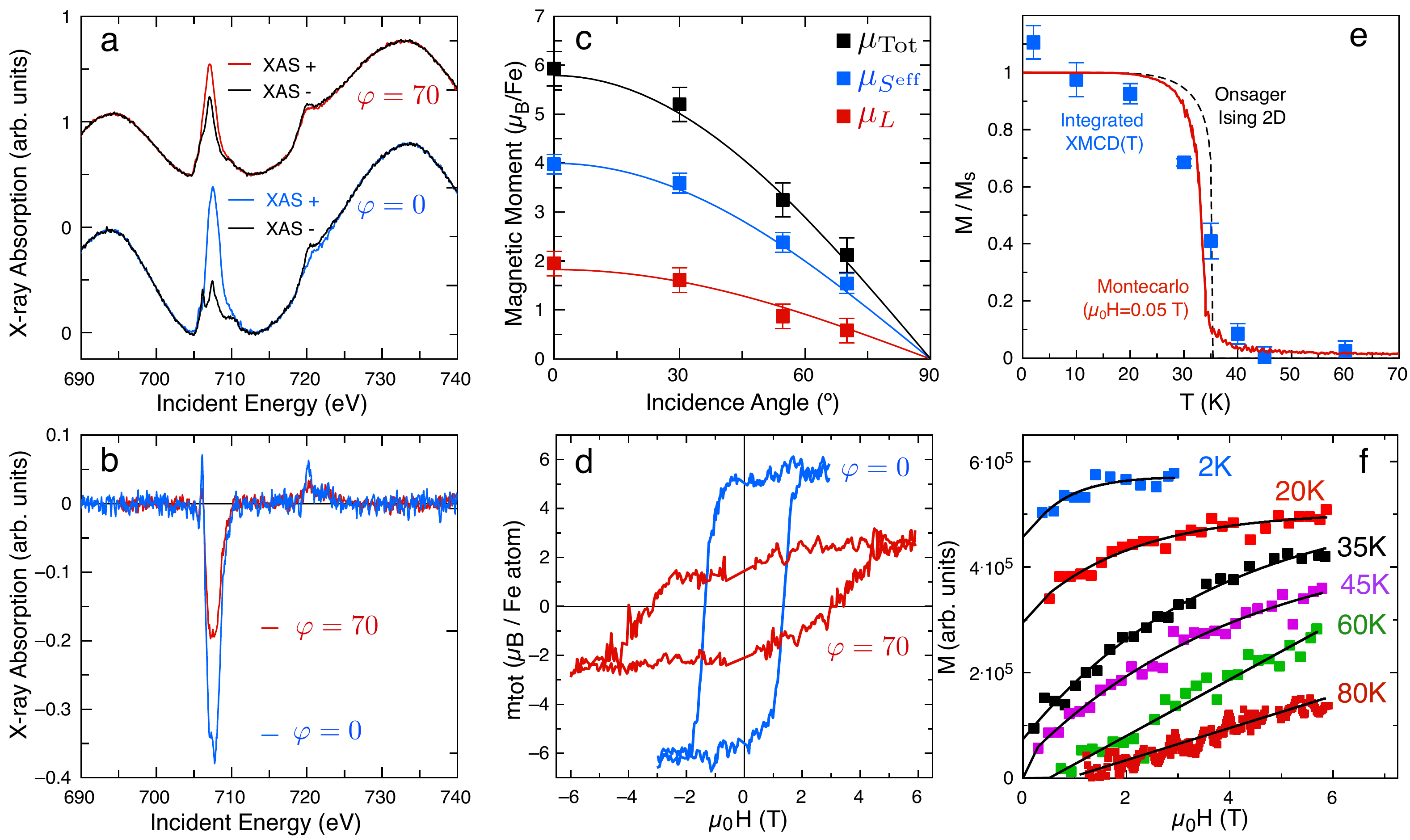}
 \end{center}
\vspace*{-3mm}
\caption[] {  
First run measurements of the Fe+DCA network on Au(111) by XMCD. These measurements were acquired using a different Au(111) substrate and with one year difference to the ones shown in Fig.~2 of the main manuscript. Data reproducibility is unquestionable judging the match of the two results. 
XAS (a) and its corresponding XMCD (b) spectra acquired with circularly right ($I^+$) and left ($I^-$) polarized X-ray beam for normal (0\de) and grazing (70\de) incidence at the $L_{2,3}$ edges of Fe. 
(c) Angular dependence of the orbital (red), effective spin (blue) and total (black) magnetic moments obtained from the sum rules in the first experimental run. Note the reproducibility of the $\cos(\varphi)$ dependence and the normal incidence values $\mu_L^{z}\approx 2$~$\mu_{\mathrm{B}}$, $\mu_S^{\mathrm{eff}_z} \approx 4$~$\mu_{\mathrm{B}}$, consistent with a Fe(II) d$^6$ high-spin (HS) configuration with $\langle L_z \rangle = 2$ and $\langle S_z \rangle = 2$.
(d) Hysteresis loop measured at the $L_3$ edge of Fe at normal (blue) and grazing (red) incidence. The coercive fields of the open loops are slightly smaller than in the main manuscript ($\approx1.5$~T for the out-of-plane and $\approx3.5$~T for in-plane). We attribute these reduction to a slightly  deffective preparation of the network on the surface on this run.
(e) Plot of the temperature dependent area enclosed in XMCD $L_3$ edge from the spectra in Fig.~S6 fitted to the Onsager analytical solution (discontinuous line) and to a Monte Carlo simulation (red line).  
(f) Magnetization isotherms at normal incidence ramping the field from 0 to 6~T. Black lines are  guide to the eyes (curved and linear below and above $T_C$, respectively), as the noise level on these measurements does not allow to perform an Arrot plot representation nor a critical analysis. Note however, that remanence is evident at and below $T_C \approx 35$~K.
 }
\label{FigS4}
\end{figure}

\begin{figure}[h]
\centering
\includegraphics[width=0.5\linewidth]{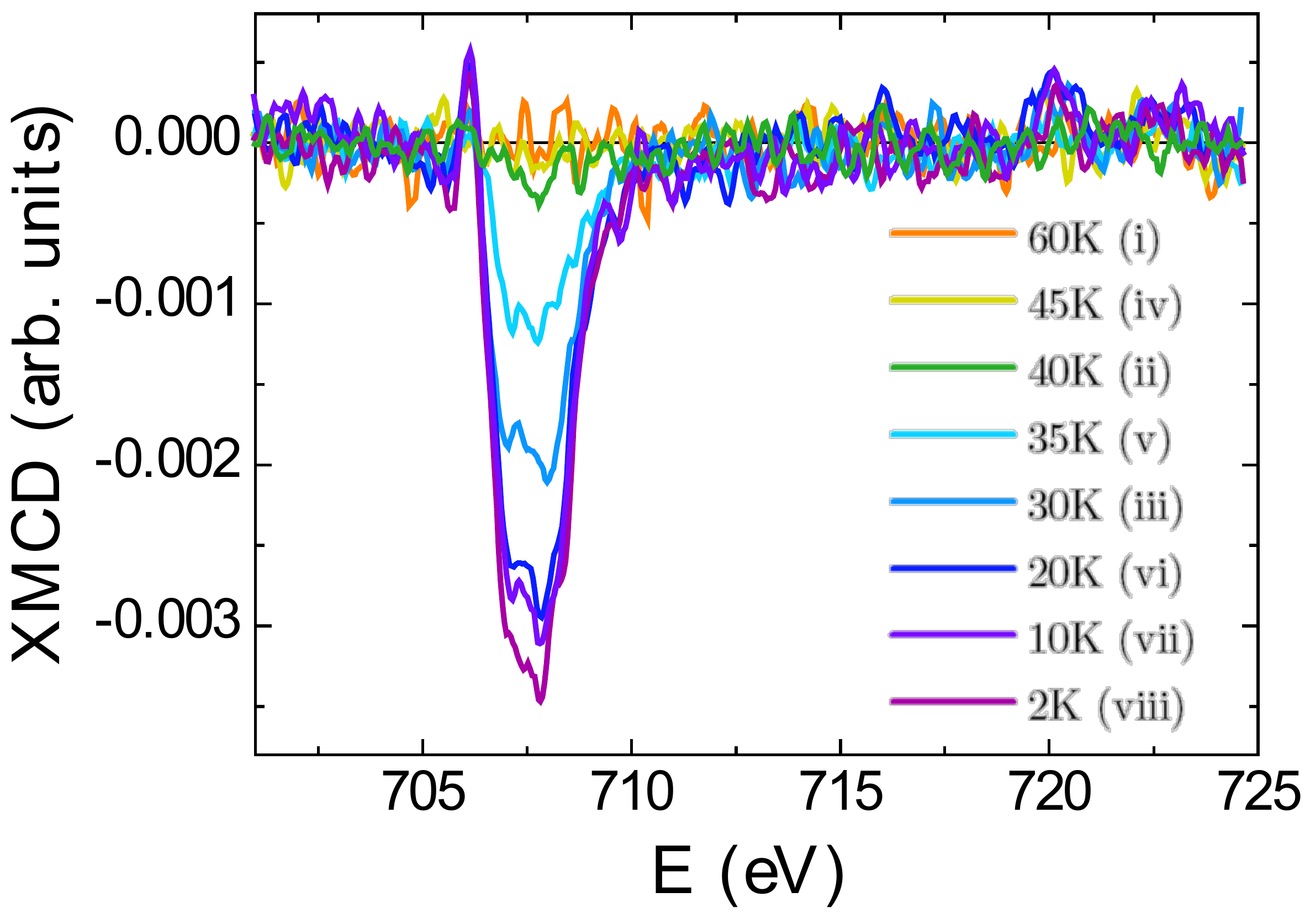}
\caption{Temperature dependence of the 2D-MOF XMCD spectra at the L$_{2,3}$ edges of Fe on the first experimental run. These XMCD spectra were sequentially recorded at normal incidence ($\varphi = 0^{\circ}$) with an external field of just H= 0.050~T. The roman numbers next to the temperatures indicate the measurement order: we started from RT and decreased the temperature until clear detection of the XMCD signal occurs at T=$30~$K. Then, the system was warmed up to T=$45~$K, where we confirmed the practical disappearance of the XMCD peaks. For a second time, we cooled down  until reaching T=$3~$K. The enclosed area of this peak was used to compose Fig.~ref{FigS4}d, which is fully compatible within the experimental uncertainty with the experimental dataset of Fig.~2e (we stress that the magnetic field used here was just 50 mT). Moreover, the time sequence of these measurements provides further evidence of the temperature reversibility of this FM system and the absence of thermal hysteresis around  $T_C \approx35$~K.
}
\label{figS6}
\end{figure}

\begin{figure}[h]
\centering
\includegraphics[width=0.85\linewidth]{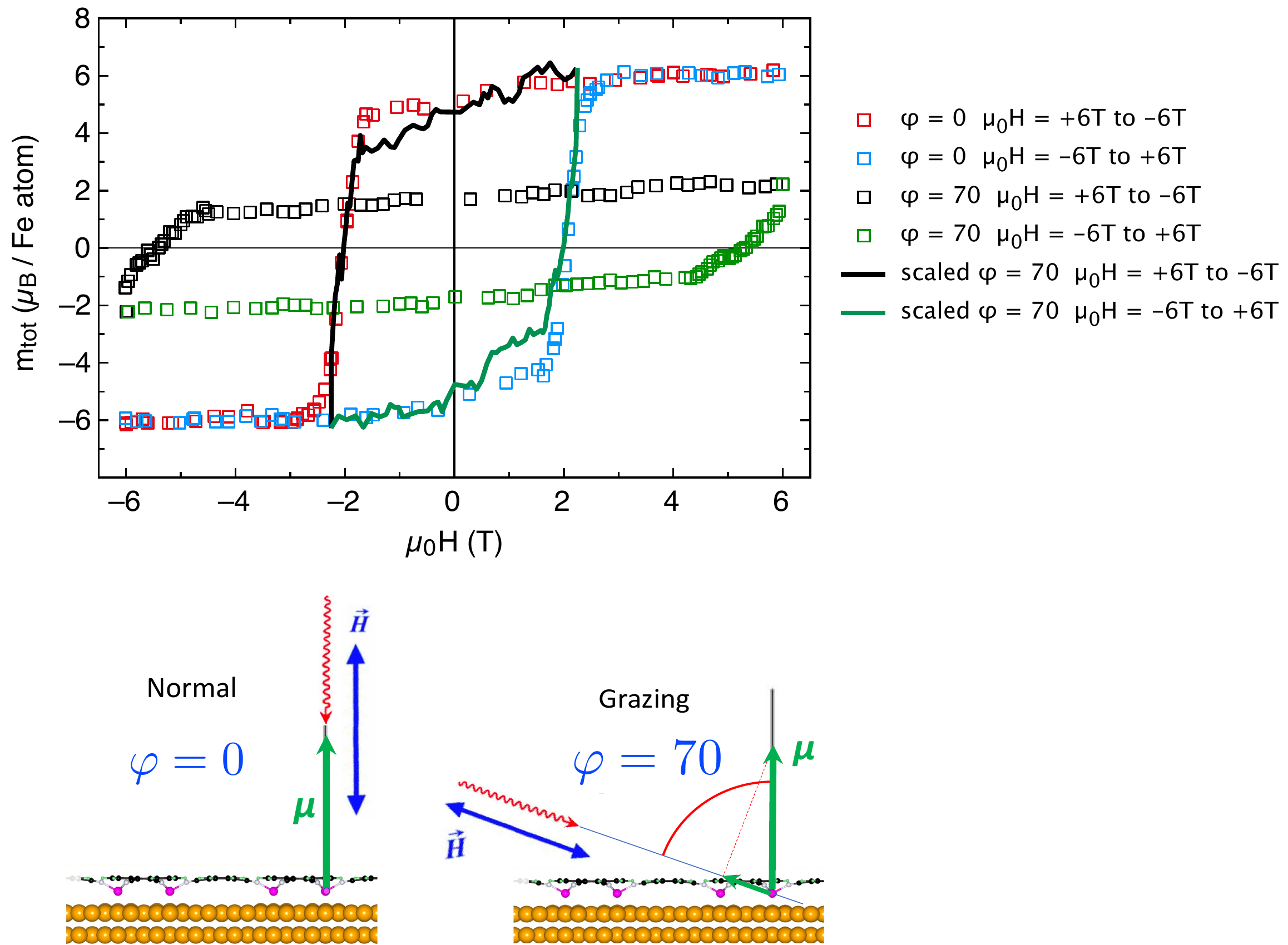}
\caption{
Hysteresis loops (open symbols) obtained at the $L_3$ edge of Fe at normal ($\varphi=0^{\circ}$) and grazing ($\varphi=70^{\circ}$) incidence (same data as in Fig.~2d of the main paper). Here the 4 branches (two for each angle incidence) have been singled out with different colors, as indicated in the legend. This figure emphasizes that the $\varphi=0^{\circ}$ loop (red and blue squares) can be calculated in very good approximation from the $\varphi=70^{\circ}$ curves (black and green squares) assuming the OOP magnetic moment (green vector in the lower figures) is not affected by more than $\pm 2\deg$ by the 6T acting on the $\varphi = 70^{\circ}$ geometry. As explained in the main text, the field along $\varphi = 0^{\circ}$ when measuring at $\varphi = 70^{\circ}$ is just the component of the applied field at $H_{0^{\circ}} = H_{70^{\circ}}\cos(70^{\circ})$ and correspondingly the  magnetization is $M_{0^{\circ}} = M_{70^{\circ}}/\cos(70^{\circ})$. For instance, if we take the $\varphi = 70^{\circ}$ dataset (green and black open squares) and calculate its projection it yields the green and black lines. Remarkably these lines  practically reproduce the $\varphi=0^{\circ}$ experimental dataset (blue and red open squares). This evidences the intense uniaxial rigidity of the magnetic moments of Fe(II) in the Fe+DCA/Au(111) system, and the strongly hard magnetic character of the 2D-MOF. 
}
\label{figS7}
\end{figure}

\begin{figure}[h]
\centering
\includegraphics[width=1.\linewidth]{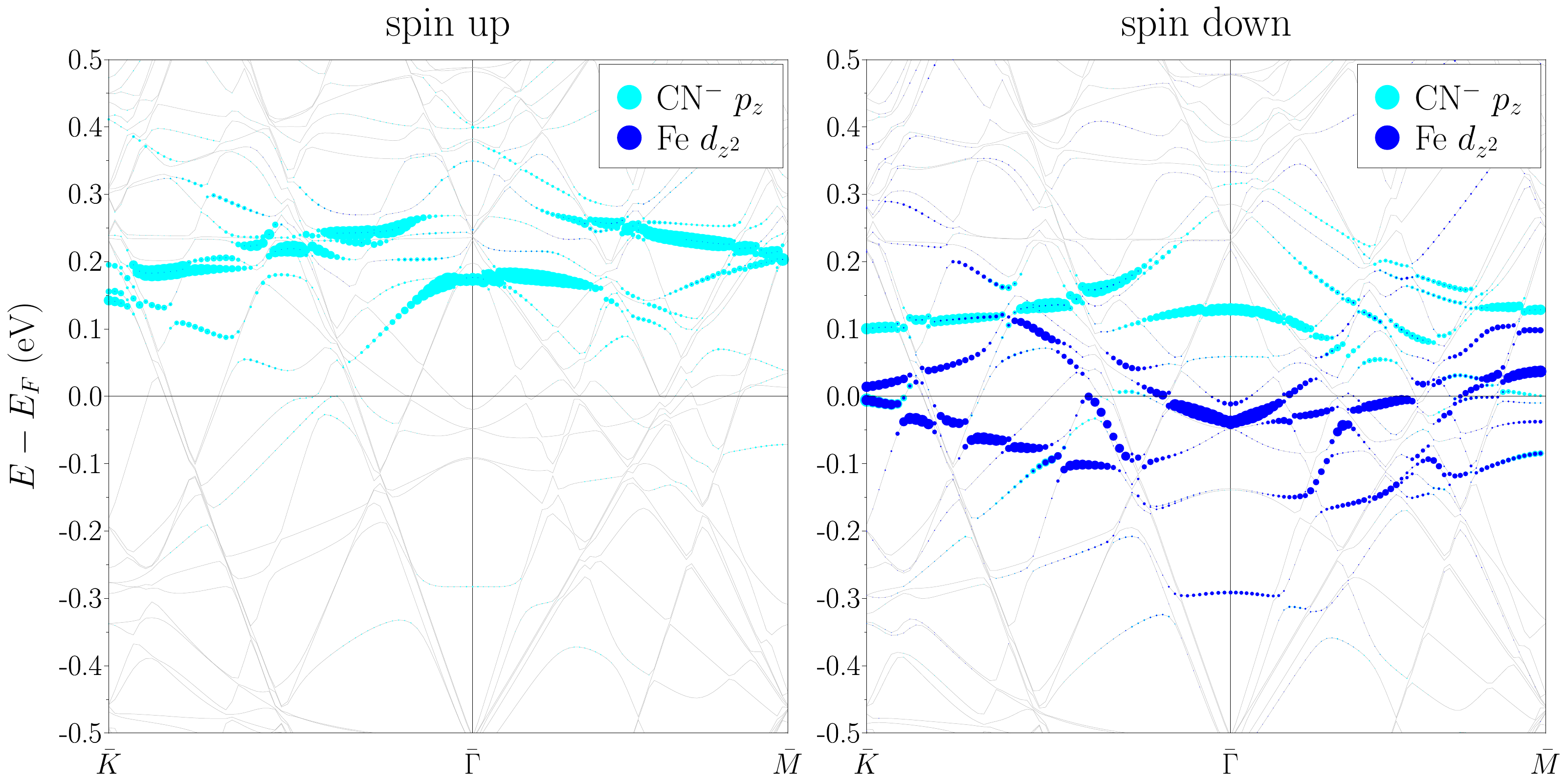}
\caption{
Calculated low-energy band structure of Fe+DCA/Au(111) using $U_{eff}=$4 eV projected onto the CN$^-$ $p_z$ (cyan) and Fe $d_{z^2}$ (blue) orbitals, which dominate in the MOF spectrum near the Fermi level. Grey lines correspond to the slab bands and are essentially the Au states. We find a strong electronic hybridization between the organic ligands and the Fe centers in comparison to the very weak overlap with the substrate. The Fe presents a spin population imbalance around the Fermi level in this $d_{z^2}$ orbital that naively agrees with their OOP ferromagnetic character.
}
\label{fig:bands}
\end{figure}

\clearpage
\break
